\documentclass[useAMS,usenatbib]{mn2e}

\usepackage{amssymb}
\usepackage{times}
\usepackage{graphicx}
\usepackage{subfigure}

\newcommand{\beq}{\begin{equation}}
\newcommand{\eeq}{\end{equation}}

\newcommand{\bfd}{\mathbf{d}}
\newcommand{\bfr}{\mathbf{r}}
\newcommand{\bfP}{\mathbf{P}}
\newcommand{\bfV}{\mathbf{V}}

\def\gs{\mathrel{\lower0.6ex\hbox{$\buildrel {\textstyle >}\over{\scriptstyle \sim}$}}}
\def\ls{\mathrel{\lower0.6ex\hbox{$\buildrel {\textstyle <}\over{\scriptstyle \sim}$}}}
\newcommand{\simgt}{\lower.5ex\hbox{$\; \buildrel > \over \sim \;$}}
\newcommand{\simlt}{\lower.5ex\hbox{$\; \buildrel < \over \sim \;$}}

\newcommand{\aap}{A\&A}
\newcommand{\apj}{ApJ}
\newcommand{\apjl}{ApJ}

\newcommand{\aj}{AJ}
\newcommand{\pasj}{PASJ}

\newcommand{\mnras}{MNRAS}

\begin{document}

\title[Triaxial lensing analysis of A1689]{Weak and strong lensing analysis of the triaxial matter distribution of A1689\thanks{Based in part on data collected at the Subaru Telescope, which is operated by the National Astronomical Society of Japan.}}
\author[Sereno \& Umetsu]{
Mauro Sereno$^{1,2}$\thanks{E-mail: mauro.sereno@polito.it (MS)} and Keiichi Umetsu$^3$
\\
$^1$Dipartimento di Fisica, Politecnico di Torino, Corso Duca degli Abruzzi 24, 10129 Torino, Italia\\
$^2$INFN, Sezione di Torino, Via Pietro Giuria 1, 10125, Torino, Italia\\
$^3$Institute of Astronomy and Astrophysics, Academia Sinica, P. O. Box 23-141, Taipei 10617, Taiwan
}


\maketitle

\begin{abstract}
Halos formed in the standard $\Lambda$ cold dark matter framework should follow an universal mass density profile and fit a well defined mass-concentration relation. Lensing analyses of clusters with a large Einstein radius seem to contradict this scenario, with the massive cluster Abell 1689 being often claimed as a notable example of a highly over-concentrated halo. Shape and orientation biases in lensing studies might be at the basis of this disagreement between theory and observations. We developed a method for a full three-dimensional analysis of strong and weak lensing data. Surface density maps estimated from lensing are de-projected to infer the actual triaxial structure of the cluster, whose mass distribution is approximated as an ellipsoidal Navarro-Frenk-White halo with arbitrary orientation. Inversion is performed under competing a priori assumptions, integrated in the method thanks to Bayesian statistics. We applied the method to A1689. Whatever the considered priors on shape and orientation, both weak and strong lensing analyses found the halo to be slightly over-concentrated but still consistent with theoretical predictions. We found some evidence for a mildly triaxial lens (minor to major axis ratio $\sim 0.5 \pm 0.2$) with the major axis orientated along the line of sight. Exploiting priors from $N$-body simulations, we found mass $M_{200} = (1.3 \pm 0.4)\times 10^{15}M_\odot$ and concentration $c_{200} =10\pm3$ for the weak lensing analysis of Subaru data, $M_{200} = (1.7 \pm 0.3)\times 10^{15}M_\odot$ and $c_{200} =6.1\pm0.9$ for the strong lensing analysis of multiple image systems, and $M_{200} = (1.3 \pm 0.2)\times 10^{15}M_\odot$ and $c_{200} =7.3\pm0.8$ for the combined weak plus strong analysis.
\end{abstract}

\begin{keywords}
galaxies: clusters: general --
        cosmology: observations  --
        gravitational lensing --
        methods: statistical --
	galaxies: clusters: individual: Abell 1689
\end{keywords}

\section{Introduction}

Clusters of galaxies are the most recent bound structures to form in a hierarchical cold dark matter model with a cosmological constant ($\Lambda$CDM). A precise determination of their mass and shape offer important clues to the understanding of the assembly process of structure in the universe. $N$-body simulations are successful in making detailed theoretical predictions on dark matter halo properties \citep{nav+al97,bul+al01,die+al04,duf+al08,pra+al11}, and lensing measurements are supposed to provide precise and model independent mass measurements. Whereas the universal Navarro-Freank-White (NFW) density profile \citep{nfw96,nav+al97} reproduces many characteristics of massive lenses, some puzzling disagreement between the  $\Lambda$CDM theoretical framework and measurements still persists. One possible conflict is the detection of extremely large Einstein radii in massive lensing clusters \citep{br+ba08,sa+re08,og+bl09,zit+al11,zit+al11b}.  Haloes should be over-concentrated to fit the data.

Concentrations of massive galaxy clusters are a crucial probe of the mean density of the universe at relatively late epochs. The concentration parameter measures the halo central density and should be related to its virial mass, with the concentration decreasing gradually with mass \citep{bul+al01}. However, cluster observations have yet to firmly confirm this correlation.

The observed concentration-mass relation for galaxy clusters has a slope consistent with what found in simulations, though the normalization factor is higher \citep{co+na07}. Lensing concentrations appear to be systematically larger than X-ray concentrations \citep{co+na07}. A similar, though less pronounced, effect is also found in simulations \citep{hen+al07,men+al10}, which show that massive lensing clusters are usually elongated along the line of sight. \citet{og+bl09} showed that the larger the Einstein radius, the larger the over-concentration problem, with clusters looking more massive and concentrated caused by the orientation bias.

The observational picture from lensing analyses is still not clear. \citet{bro+al08} derived lens distortion and magnification of four nearly relaxed high-mass clusters, inferring significantly high concentrations. The analysis of some clusters with very large Einstein radii \citep{zit+al09,zit+al11} also showed a strong over-concentration problem. \citet{ogu+al09} found that the data from a sample of ten clusters with strong and weak lensing features were highly inconsistent with the predicted concentration parameters, even including a 50 per cent enhancement to account for the lensing bias. Strong lensing features in the MACS cluster sample \citep{zit+al11,men+al11} confirmed the over-concentration problem.

On the other hand, out of 30 X-ray clusters with significant weak lensing signal, \citet{oka+al10} found that the 19 clusters that were well fitted by a NFW profile showed a correlation in the $c(M)$ relation which is marginally compatible with predictions for both slope and normalization. Finally, weak lensing analyses of stacked clusters of lesser mass do not exhibit the high concentration problem \citep{joh+al07,man+al08}, in agreement with the findings of \citet{og+bl09}.

Several effects can play a role. Strong lensing clusters tend to preferentially sample the high-mass end of the cluster mass function \citep{co+na07}. Lensing concentrations can be also inflated due to substructures close to the line of sight.  Furthermore, without adequate colour information, the weak-lensing signal can be diluted particularly toward the cluster centre by the inclusion of unlensed cluster members in the definition of background galaxy samples \citep{bro+al05b}, leading to an underestimation
of the concentration \citep{lim+al07, ume+bro08} and of the central cluster mass examined at high overdensities \citep{oka+al10} 

Disagreement between theory and observation might be explained by some orientation and shape biases. Triaxial halos can be much more efficient lenses than their more spherical counterparts \citep{og+bl09} with the strongest lenses in the universe expected to be a highly biased population preferentially orientated along the line of sight and with high levels of triaxiality. On the other hand, \citet{ham+al09} carried out a spectroscopic follow-up survey of weak-lensing selected cluster candidates of \citet{miy+al02}, and found no evidence of lensing selection bias in mass estimates compared to X-ray and optical techniques.

Only a few works have tried to infer mass and concentration from lensing data without assuming spherical symmetry. \citet{ogu+al05} studied the mass profile of A1689 exploiting both strong and weak lensing measurements. They showed that the intrinsic model parameters are weakly constrained (mainly because of the lack of information along the line of sight), and that A1689 lensing observations might be compatible with predictions assuming a major axis of the halo aligned with the line of sight. Triaxiality issues were addressed by \citet{cor+al09}, who put weak lensing constraints on three strong lensing clusters exploiting Bayesian statistical methods. They found that the large errors that accompany triaxial parameter estimates can make observations compatible, even if marginally, with theoretical predictions. In fact, neglecting halo triaxiality can lead to over- and under-estimates of up to 50 per cent and a factor of 2 in halo mass and concentration, respectively \citep{co+ki07}. \citet{ser+al10} investigated a sample of 10 strong lensing clusters considering which intrinsic shape and orientation the lensing halos should have to account for both theoretical predictions and observations. They found that nearly one half of the clusters seemed to be composed of outliers of the mass-concentration relation, whereas the second half supported expectations of $N$-body simulations which prefer mildly triaxial lensing clusters with a strong orientation bias. The strong lensing analysis of AC~114 also supported that mapping effects can play a role in the estimate of the concentration \citep{ser+al09}.  

Here we apply a novel lensing inversion method to infer intrinsic cluster parameters. We exploit both weak and strong lensing data through a full two-dimensional (2D) analysis and then deproject the lensing derived surface mass density with a Bayesian approach. The method is applied to Abell 1689 (A1689), a very massive cluster at redshift $z=0.183$ with one of the largest Einstein radius observed to date, $\sim 45\arcsec$ for a source redshift  $z_\mathrm{s}=1$ \citep{bro+al05,lim+al07}. Despite being thoroughly investigated it is still unclear if and how much this cluster is over-concentrated. Independent strong lensing analyses of the inner regions provided a consistent picture of the mass distribution in the inner $\ls 300~\mathrm{kpc}$ \citep{bro+al05,hal+al06,lim+al07,coe+al10},  but results are not fully in agreement on the larger virial scale. Different weak lensing analyses of either Subaru \citep{ume+bro08,ume+al09}, ESO/MPG \citep{cor+al09}, or CFH12K \citep{lim+al07} suggest somewhat different degrees of concentration, $c_{200}$. Subaru  and CFHT results are marginally consistent: $c_{200}=10.1\pm 0.8 (\pm 2.2$ systematic) in \citet{ume+bro08}; $c_{200}=7.6\pm 1.6$ in \citet{lim+al07}. The broad conclusion is that weak lensing investigations point out to a more concentrated cluster than strong lensing \citep{coe+al10}, but the quantitative assessment of such tension is still to be firmly ascertained.  

One way to possibly solve the overconcentration problem in A1689 is to properly consider the triaxial structure of the cluster. \citet{ogu+al05} first noted that measured concentration in lensing analyses of A1689 would be marginally in agreement with theoretical prediction if the major axis of the cluster is aligned with the line of sight. \citet{def+al05} and \citet{ser+al06} combined X-ray and Sunyaev-Zel'dovich (SZ) observations to infer the elongation of the hot intra-cluster medium without assuming hydrostatic equilibrium and found that A1689 was quite spherical within the errors. A triaxial weak lensing analysis performed in \citet{cor+al09} revealed that realistic intrinsic parameter estimates in A1689 must be accompanied by larger errors due to our limited knowledge of the halo shape and orientation. 

The hypothesis of hydrostatic equilibrium in A1689 was exploited by \citet{lem+al08}, who combined {\it Chandra} X-ray brightness measurements and joint strong/weak lensing measurements of the projected  cluster mass profile by \citet{bro+al05} to derive a temperature profile that is consistent with hydrostatic equilibrium. It was found that the resulting equilibrium temperature exceeds the observed temperature by 30 per cent at all radii. \citet{lem+al08} suggested that the existence of significant cool and dense gas components is the source of this temperature discrepancy \citep{kaw+al07}.   

On the basis of a detailed X-ray analysis of deep {\it Chandra} observations, \citet{pen+al09} showed that the hydrostatic mass within the central 200\,kpc$\,h^{-1} $region of A1689 is about 30--50 per cent lower than lensing-based mass estimates obtained under the assumption of spherical symmetry. They also showed that it is unlikely that dense and cool gas clumps alone can cause such a strong bias in the X-ray temperature determinations, since the space filling factor for such components should be 70--90 per cent. They suggested instead  that a prolate ellipsoidal configuration, aligned with the line of sight, could solve the central mass discrepancy assuming an axis ratio of 0.6 for the gas distribution, which however overestimates the projected total mass at large radii by $\sim 40$ per cent. \citet{mor+al11} confirmed that if the halo of A1689 is very elongated along the line of sight, strong lensing and X-ray observations are compatible with theoretical predictions on the $c(M)$ relation. 

The level of hydrostatic equilibrium in A1689 has been recently explored by \citet{kaw+al10} and \citet{mol+al10}.  Deep mosaic {\it Suzaku} X-ray observations by \citet{kaw+al10}, covering the entire cluster, revealed anisotropic temperature and entropy distributions in cluster outskirts correlated with large-scale structure of galaxies. Three out of four quadrants of the mosaic image, in contact with low density void environments, have low gas temperatures and entropies, deviating from hydrostatic equilibrium. \citet{kaw+al10} showed that the thermal gas pressure within $\ls 1.3\,$Mpc  (about half the virial radius)  is at most about half of the equilibrium pressure to balance the gravity predicted by gravitational lensing under the hypothesis of spherical symmetry, and $\sim 30-40$ per cent around the virial radius. They suggested that additional sources of pressure support, such as bulk and/or turbulent motions, are required in the central and outskirt regions.

\citet{mol+al10} challenged the assumption of hydrostatic equilibrium in the core region of A1689. By analysing a sample of high-mass clusters with $M =(1-2)\times 10^{15} M_\odot$ drawn from high-resolution cosmological simulations, they found in the core region of relaxed clusters, selected based on their smooth spherically averaged density profile and no sign of recent major merger events, significant non-thermal pressure support of $\sim 20$ per cent from subsonic random gas motions within 1--10 per cent of the virial radius \citep{you+bry07,lau+al09}. They suggested that discrepancies might be due in part to non-thermal pressure support.

Here, we reconsider the estimate of the concentration in A1689 from lensing analyses renouncing the spherical hypothesis. The paper is organised as follows. In Sec.~\ref{sec_nfw}, we review the properties of the three-dimensional NFW profile and how it projects on the plane of the sky. Theoretical predictions derived in $N$-body simulations are discussed in Sec.~\ref{sec_theo}. Fitting results to strong and weak lensing data can be found in Sec.~\ref{sec_lens}. The lensing inversion is discussed in Sec.~\ref{sec_inve}, whereas Sec.~\ref{sec_resu} focuses on results. Section~\ref{sec_conc} is devoted to the conclusions. Throughout the paper, we assume a flat $\Lambda$CDM cosmology with density parameters $\Omega_\mathrm{M}=0.3$, $\Omega_{\Lambda}=0.7$ and Hubble constant $H_0=100h~\mathrm{km~s}^{-1}\mathrm{Mpc}^{-1}$, $h=0.7$. At the A1689 distance, $1\arcsec$ corresponds to $3.08~\mathrm{kpc}$.

\section{Basics on triaxial NFW halo}
\label{sec_nfw}

High resolution $N$-body simulations have shown that the density distribution of dark matter halos is successfully described as a NFW density profile \citep{nfw96,nav+al97}, whose 3D distribution follows
\begin{equation}
\label{nfw1}
	\rho_\mathrm{NFW}=\frac{\rho_\mathrm{s}}{(r/r_\mathrm{s})(1+r/r_\mathrm{s})^2},
\end{equation}
where $\rho_\mathrm{s}$ is the characteristic density and $r_\mathrm{s}$ is the characteristic length scale. Halos are expected to be aspherical and their profiles can be accurately described by concentric triaxial ellipsoids with aligned axes \citep{ji+su02}. The spherical radius $r$ has to be replaced with an ellipsoidal radial variable. The minor (intermediate) to major axial ratio is denoted as $q_1$ ($q_2$) with $0< q_1 \le q_2\le 1$; we also use the inverse ratios, $0< e_i =1/q_i \ge 1$.

Different definitions of parameters for spherically averaged profiles can play a role when comparing observations to predictions \citep{br+ba08}. Here, we follow \citet{co+ki07}, who generalised NFW parameter definitions to ellipsoidal halos. The radius $r_{200}$ is such that the mean density contained within an ellipsoid of semi-major axis $r_{200}$ is $\Delta= 200$ times the critical density at the halo redshift, $\rho_\mathrm{cr}$; the corresponding concentration is $c_{200} \equiv r_{200}/ r_\mathrm{s}$. $M_{200}$ is the mass within the ellipsoid of semi-major axis $r_{200}$, $M_{200}=(800\pi/3)q_1q_2 r_{200}^3 \rho_\mathrm{cr}$. Such defined $c_{200}$ and $M_{200}$ have small deviations with respect to the parameters computed fitting spherically averaged density profiles, as done in  $N$-body simulations. The only caveat is that the spherical mass obtained in simulations is significantly less than the ellipsoidal $M_{200}$ for extreme axial ratios \citep{co+ki07}. However, since the dependence of the concentration on the mass is quite weak, see Eq.~(\ref{nbod2}), effects on our analysis are negligible.

Three Euler's angles, $\theta, \varphi$ and $\psi$, relate the intrinsic to the observer's coordinate system; the angle $\theta$ quantifies the inclination of the major axis with respect to the line of sight. When viewed from an arbitrary direction, quantities constant on similar ellipsoids project themselves on similar ellipses~\citep{sta77}.

\subsection{Projected density}

In gravitational lensing studies, the projected map of interest is the surface mass density, expressed in terms of the convergence $\kappa$, i.e., in units of the critical surface mass density for lensing, $\Sigma_\mathrm{cr}=(c^2\,D_\mathrm{s})/(4\pi G\,D_\mathrm{d}\,D_\mathrm{ds})$, where $D_\mathrm{s}$, $D_\mathrm{d}$ and $D_\mathrm{ds}$ are the source, the lens and the lens-source angular diameter distances respectively. The projected surface density has the same functional form as a spherically symmetric halo. The axial ratio of the major to the minor axis of the observed projected isophotes, $e_{\rm P}(\geq 1)$, can be written as \citep{bin80},
\begin{equation}
\label{eq:tri4e}
e_{\rm P}= \sqrt{ \frac{j+l + \sqrt{(j-l)^2+4 k^2 } }{j+l
-\sqrt{(j-l)^2+4 k^2 }} },
\end{equation}
where  $j, k$ and $l$ are defined as
\begin{eqnarray}
j & = &  e_1^2 e_2^2 \sin^2 \theta + e_1^2 \cos^2 \theta \cos^2 \varphi + e_2^2 \cos^2 \theta \sin^2 \varphi ,  \label{eq:tri4a} \\
k & = &  (e_1^2 - e_2^2) \sin \varphi \cos \varphi \cos \theta ,  \label{eq:tri4b}  \\
l & = &  e_1^2 \sin^2 \varphi + e_2^2 \cos^2 \varphi . \label{eq:tri4c}
\end{eqnarray}
We also use the ellipticity $\epsilon = 1-1/e_\mathrm{P}$.

The projected NFW density profile can be described in terms of the strength of the lens $k_\mathrm{s}$ and of the projected length scale $r_\mathrm{sP}$, i.e., the two parameters directly inferred by fitting lensing maps \citep{ser+al09}. The observed scale length $r_\mathrm{sP}$ is the projection on the plane of the sky of the cluster intrinsic length \citep{sta77,ser07},
\begin{equation}
\label{eq:tri6}
r_{\rm sP} = r_{\rm s} \left( \frac{e_\Delta}{\sqrt{f}} \right),
\end{equation}
where $f$ is a function of the cluster shape and orientation,
\begin{equation}
\label{eq:tri3}
f = e_1^2 \sin^2 \theta \sin^2 \varphi + e_2^2 \sin^2 \theta \cos^2 \varphi + \cos^2 \theta ;
\end{equation}
the parameter $e_\Delta$ quantifies the elongation of the triaxial ellipsoid along the line of sight,
\beq
e_\Delta = \left( \frac{e_\mathrm{P}}{e_1 e_2}\right)^{1/2} f^{3/4};
\eeq
$e_\Delta$ represents the ratio between the major axis of the projected ellipse in the plane of the sky and the size of the ellipsoid along the line of sight. When $e_\Delta <1$, the cluster is more elongated along the line of sight than wide in the plane of the sky, i.e., the smaller the $e_\Delta$ parameter, the larger the elongation along the line of sight.

The relation between the projected parameters, $k_\mathrm{s}$ and $r_\mathrm{sP}$, and the intrinsic parameters can be written as \citep{ser+al09},
\beq
\label{nfw1b}
\Sigma_\mathrm{cr} \times k_\mathrm{s} = \frac{f_\mathrm{geo}}{\sqrt{e_\mathrm{P}}}\rho_\mathrm{s} r_\mathrm{sP},
\eeq
where as usual $\rho_\mathrm{s} = \delta_c \rho_\mathrm{cr}$,
\beq
\delta_c = \frac{200}{3}\frac{c_{200}}{\ln (1+c_{200})-c_{200}/(1+c_{200})},
\eeq 
and $f_\mathrm{geo} \equiv \sqrt{e_\mathrm{P}}/e_\Delta$. The mass $M_{200}$ can be expressed as
\beq
\label{nfw2}
M_{200}= \frac{4\pi}{3}\times 200 \rho_\mathrm{cr} \times (c_{200} r_\mathrm{sP})^3 \frac{f_\mathrm{geo}}{e_\mathrm{P}^{3/2}}.
\eeq

\section{Theoretical predictions}
\label{sec_theo}

$N$-body simulations \citep{net+al07,mac+al08,gao+al08,duf+al08,pra+al11} have provided a quite detailed picture of the expected properties of dark matter halos. Results may depend on the overall normalization of the power spectrum, the mass resolution and the simulation volume \citep{pra+al11}. The dependence of halo concentration on  mass and redshift can be adequately described by a power law,
\beq
\label{nbod1}
c =A(M/M_\mathrm{pivot})^B(1+z)^C.
\eeq
Here, we follow \citet{duf+al08}, who used the cosmological parameters from WMAP5 and found $\{A,B,C\}=\{ 5.71 \pm0.12, -.084 \pm 0.006, -0.47\pm0.04\}$ for a pivotal mass $M_\mathrm{pivot}=2\times10^{12}M_\odot/h$ in the redshift range $0-2$ for their full sample of clusters. The scatter in the concentration about the median $c(M)$ relation has been shown to be lognormal,
\beq
\label{nbod2}
p(\ln c | M)=\frac{1}{\sigma\sqrt{2\pi}}\exp \left[ -\frac{1}{2} \left(  \frac{\ln c - \ln c(M)}{\sigma}\right) \right],
\eeq
with a dispersion $\sigma (\log_{10} c_{200})=0.15$ for a full sample of clusters \citep{duf+al08}.

\citet{ji+su02} investigated the probability distribution of intrinsic axial ratios and proposed an universal approximating formula for the distribution of minor to major axis ratios,
\beq
\label{nbod3}
p(q_1) \propto \exp \left[ -\frac{(q_1-q_\mu/r_{q_1})^2}{2\sigma_\mathrm{s}^2}\right]
\eeq
where $q_\mu=0.54$, $\sigma_\mathrm{s}=0.113$ and 
\beq
r_{q_1} = (M_\mathrm{vir}/M_*)^{0.07 \Omega_M(z)^{0.7}},
\eeq
with $M_*$ the characteristic nonlinear mass at redshift $z$ and $M_\mathrm{vir}$ the virial mass. The conditional probability for $q_2$ goes as 
\beq
\label{nbod4}
p(q_1/q_2|q_1)=\frac{3}{2(1-r_\mathrm{min})}\left[ 1-\frac{2q_1/q_2-1-r_\mathrm{min}}{1-r_\mathrm{min}}\right]
\eeq
for $q_1/q_2 \geq r_\mathrm{min} \equiv \max[q_1,0.5]$, whereas is null otherwise. The lensing population has nearly the same triaxialily distribution as the total cluster population \citep{hen+al07}.

For comparison we also consider a flat distribution for the axial ratio in the range $q_\mathrm{min}<q_1 \le 1$ and $q_1 \le q_2 \le 1$. Probabilities are defined such that the marginalized probability $p(q_1)=const.$ and the conditional probability $p(q_2|q_1)=const.$ The probabilities can then be expressed as
\beq
\label{flat1}
p(q_1) =1/(1-q_\mathrm{min})
\eeq 
for $q_\mathrm{min}<q_1 \le 1$ and zero otherwise;
\beq
\label{flat2}
p(q_2|q_1) = (1-q_1)^{-1}
\eeq 
for $q_2 \ge q_1$ and zero otherwise. The resulting probability for $q_2$ is then 
\beq
p(q_2) =\frac{1}{1-q_\mathrm{min}} \ln \left(\frac{1-q_\mathrm{min}}{1-q_2}\right).
\eeq 
A flat distribution allows also for very triaxial clusters ($q_1 \ls q_2 \ll1$), which are preferentially excluded by $N$-body simulations.

Finally, semi-analytical \citep{og+bl09} and numerical \citep{hen+al07} investigations showed a large tendency for lensing clusters to be aligned with the line of sight. Such condition can be expressed as \citep{cor+al09}
\beq
\label{nbod5}
p(\cos \theta) \propto \exp \left[-\frac{(\cos \theta -1)^2}{2\sigma_\theta^2}\right],
\eeq
with $\sigma_\theta=0.115$. For comparison, we also consider a population of randomly oriented clusters with
\beq
\label{flat3}
p(\cos \theta) = 1
\eeq 
for $0 \le \cos \theta \le 1$.

\section{Lensing analysis}
\label{sec_lens}

In this section, we fit strong and weak lensing data with a projected ellipsoidal NFW halo.

\begin{table*}
\centering
\begin{tabular}[c]{lr@{$\,\pm\,$}lr@{$\,\pm\,$}lr@{$\,\pm\,$}lr@{$\,\pm\,$}lr@{$\,\pm\,$}lr@{$\,\pm\,$}l}
        \hline
        \noalign{\smallskip}
        Component & \multicolumn{2}{c}{mass scale} & \multicolumn{2}{c}{$\theta_{1,0}$} & \multicolumn{2}{c}{$\theta_{2,0}$}& \multicolumn{2}{c}{$\epsilon$}&\multicolumn{2}{c}{$\theta_\epsilon$} & \multicolumn{2}{c}{length scale}    \\
        \noalign{\smallskip}
        & \multicolumn{2}{c}{} & \multicolumn{2}{c}{($\arcsec$)} &\multicolumn{2}{c}{($\arcsec$)} & \multicolumn{2}{c}{} &
\multicolumn{2}{c}{($\deg$) }&\multicolumn{2}{c}{($\mathrm{kpc}$)}  \\
        \noalign{\smallskip}
        \hline
          \multicolumn{13}{c}{Strong Lensing} \\
       & \multicolumn{2}{c}{$k_\mathrm{s}$} & \multicolumn{8}{c}{}  &\multicolumn{2}{c}{$r_\mathrm{sP}$}   \\
        NFW & $0.35(0.37)$&$0.04$& $0.0(0.4)$&$0.6$ & $-4.6(-4.4)$&$0.6$ & $0.29(0.27)$&$0.04$ & $-26(-24)$&$5$ & $400(370)$&$70$   \\
        &	\multicolumn{2}{c}{$b~(\arcsec)$} &	\multicolumn{8}{c}{}&\multicolumn{2}{c}{$r_\mathrm{cP}$}  \\
        \noalign{\smallskip}
       p-Jaffe & $15(13)$&$5$& $-80(-70)$&$30$ & $80(70)$&$40$ & $0.88(0.88)$&$0.06$ & $-142(-143)$&$2$ & $43(38)$&$23$   \\
       \hline
       \multicolumn{13}{c}{Weak Lensing} \\
         & \multicolumn{2}{c}{$k_\mathrm{s}$} & \multicolumn{8}{c}{}  &\multicolumn{2}{c}{$r_\mathrm{sP}$}   \\
        NFW & $0.62(0.53)$&$0.24$& $-6(-6)$&$7$ & $-5(-7)$&$5$ & $0.17(0.21)$&$0.12$ & $30(30)$&$30$ & $210(230)$&$70$   \\
        \hline
       \multicolumn{13}{c}{Weak and Strong Lensing} \\
         & \multicolumn{2}{c}{$k_\mathrm{s}$} & \multicolumn{8}{c}{}  &\multicolumn{2}{c}{$r_\mathrm{sP}$}   \\
        NFW & $0.41(0.42)$&$0.03$& $-0.1(0.1)$&$0.6$ & $-4.6(-4.5)$&$0.6$ & $0.30(0.29)$&$0.04$ & $-22(-21)$&$4$ & $310(300)$&$30$   \\
\hline
\end{tabular}
\caption{
Projected model parameters inferred with the lensing analysis. The orientation angle $\theta_\epsilon$ is measured north over east and $r_\mathrm{sP}$ ($r_\mathrm{cP}$) is the projected length scale (core radius) for the NFW (p-Jaffe) profile. Convergence is normalised to a reference source redshift ($z_\mathrm{s}=2.0$). The truncation radius of the p-Jaffe component was kept fixed to 500~kpc. For the WL plus SL analysis, we report only the results for the main halo. Central location and dispersion of the PDF are computed as biweight estimators \citep{bee+al90}. Maximum likelihood values are reported in parentheses.
}
\label{tab_fit_nfw_sl_wl}
\end{table*}

\begin{figure*}
\begin{center}
$
\begin{array}{c@{\hspace{.1in}}c@{\hspace{.1in}}c@{\hspace{.1in}}c}
\includegraphics[width=5cm]{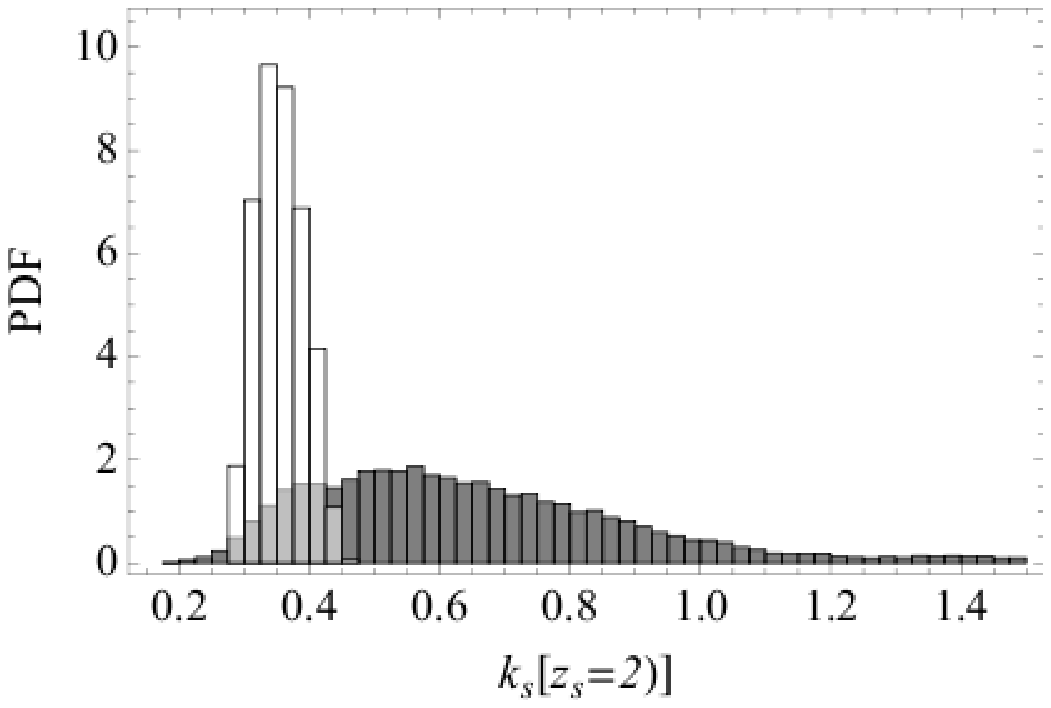} &
\includegraphics[width=5cm]{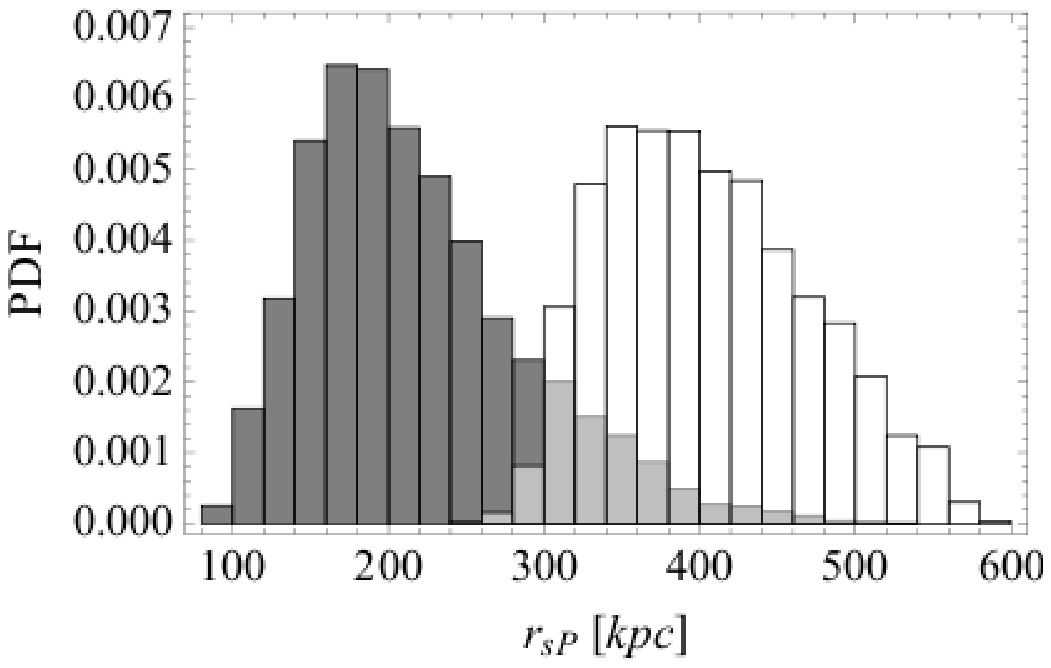}&
\includegraphics[width=5cm]{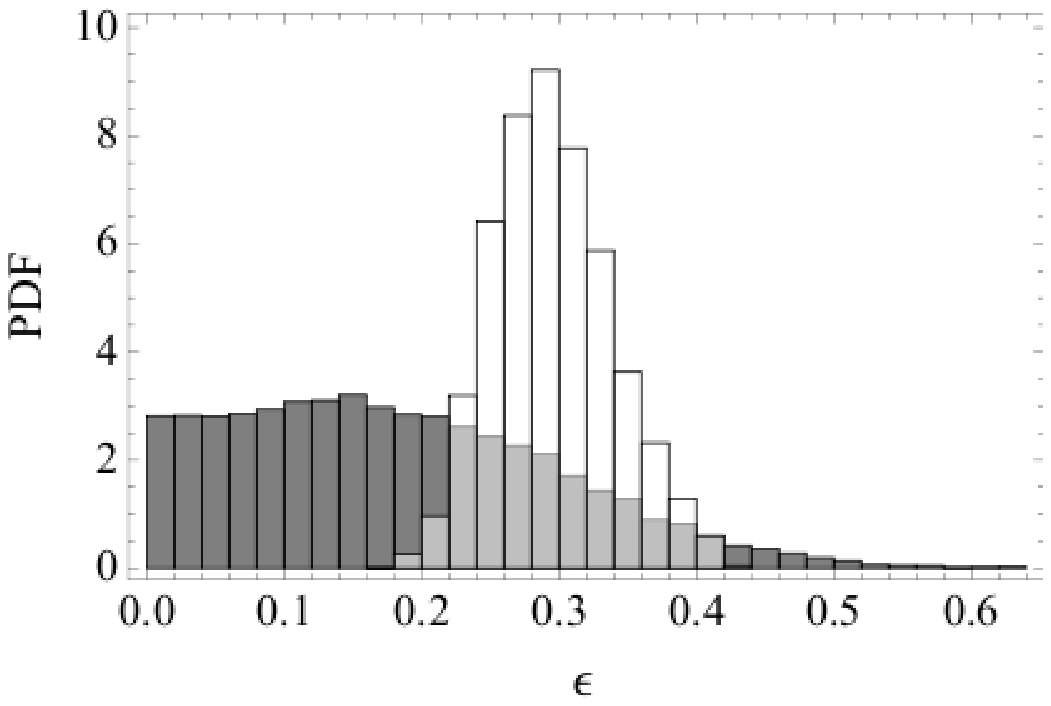}
\end{array}
$
\end{center}
\caption{Marginalised posterior probability density functions for the projected parameters of the NFW halo. Panels from the left to the right are for the binned PDF of $k_\mathrm{s}$, $r_{sP}$, and $\epsilon$, respectively. Shadowed and white bins are for the weak and strong lensing results, respectively.}
\label{fig_histo_ks_sl_wl_redsex}
\end{figure*}

\subsection{Weak lensing}
\label{sec_like_wl}

A1689 has been object of very detailed weak lensing (WL) studies. Here, we rely on previous work by \citet{ume+bro08} and \citet{ume+al09} on wide-field Subaru data. They obtained the two-dimensional convergence map reconstructed from a regularised maximum-likelihood combination of the joint measurements of the weak-lensing distortion and magnification bias effects on background red galaxies, the combination of which can break the mass-sheet degeneracy inherent in all reconstruction methods based on distortion information alone \citep{bro+al95}.  The map is given on $21\times 17(=357)$ grid pixels covering a field of $\sim 30\arcmin \times 24\arcmin$. Such a map probes the cluster halo up to and beyond the virial radius ($\sim 16\arcmin$) with the accuracy of the pixel size ($=1.4\arcmin$). The map we considered was obtained with a pure WL (shear plus magnification) reconstruction method, described as ``2D MEM" in \citet{ume+bro08}.

A full two-dimensional weak lensing analysis is crucial for comparison with theoretical predictions \citep{ogu+al10}. Given the measured convergence map $k_\mathrm{obs}$, the weak lensing $\chi^2_\mathrm{WL}$ function can be expressed as \citep{ogu+al05}, 
\beq
\label{like_wl}
\chi^2_\mathrm{WL}=\sum_{i,j}\left[ k_\mathrm{obs}(\bfr_i)- k(\bfr_i) \right] \left( V^{-1}\right)_{i,j} \left[ k_\mathrm{obs}(\bfr_j)- k(\bfr_j) \right] 
\eeq
where $\bfV^{-1}$ is the inverse of the pixel-pixel covariance matrix. The corresponding likelihood is ${\cal{L}}_\mathrm{WL} \propto \exp(-\chi^2_\mathrm{WL}/2)$. The $\chi^2 (\simeq 319)$ for the best fitting model (found with a downhill simplex algorithm) is very similar to what found in \citet{ume+bro08} or \citet{ume+al09}. The slight improvement is due to the use of an elliptical model. We run four Markov chains of 50000 samples each after an early parameter diagonalization. Chain convergence was reached. For each series of chains run in the present paper, we checked that the standard var(chain mean)/mean(chain var) indicator was less than 1.2. Results are summarised in Table~\ref{tab_fit_nfw_sl_wl}.

\subsection{Strong lensing analysis}
\label{sec_like_sl}

\begin{figure}
       \resizebox{\hsize}{!}{\includegraphics{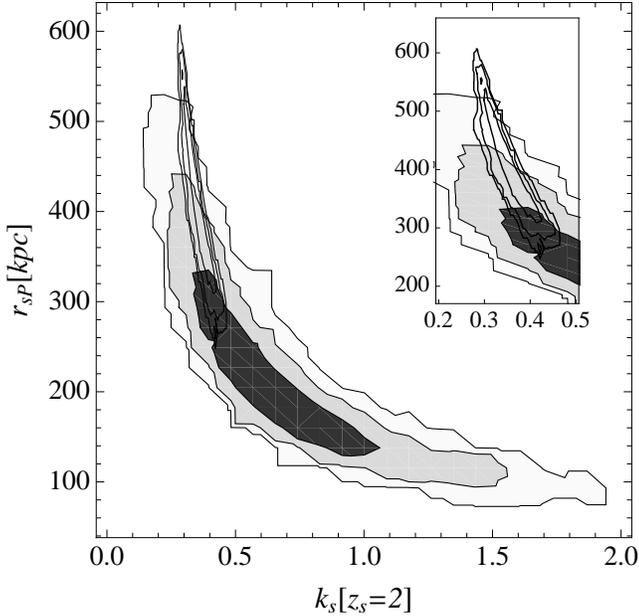}}
       \caption{Contour plot of the marginalised PDF for the strength $k_\mathrm{s}$ and the projected radius $r_{sP}$. Contours are plotted at fraction values $\exp (-2.3/2)$, $\exp(-6.17/2)$, and $\exp(-11.8/2)$ of the maximum, which denote confidence limit region of 1, 2 and $3\sigma$ in a maximum likelihood framework, respectively. The shadowed regions are for the WL results, the thick contours for the SL. The inset zooms in for the SL region.}
	\label{fig_ks_rsP_sl_wl_redsex}
\end{figure}

The HST observations of Abell 1689 are among the deepest to date for any galaxy cluster. The cluster has been observed in four filters ($g_{475}$ , $r_{625}$ , $i_{775}$, and $z_{850}$ passbands) with ACS during 20 orbits of HST and eight more bands from the ground \citep{bro+al05}. The core of A1689 is very rich in detected lensing features: 135 candidate images of 42 galaxies were identified \citep{lim+al07,coe+al10}. We considered only systems with confirmed spectroscopic redshifts, which left us with 80 confirmed images in 24 systems\footnote{Systems 1, 2, 4, 5, 6, 7, 10, 11, 12, 14, 15, 17, 18, 19, 22,  24, 29, 30, 31, 32, 33, 35,  36, 40 according to the notation in \citet{lim+al07}.}. 

We employed a parametric lensing analysis method based on the {\it gravlens} kernel evaluation of $\chi^2_\mathrm{SL}$ values \citep{kee01b,kee01a}. As usual, strong lensing (SL) likelihood values were computed by comparing observed to predicted image positions.

Since we are interested in constraining the total mass distribution, we tried to keep the model as simple as possible. For an ideal comparison with theoretical predictions, we should model the total matter distribution in the cluster as a single NFW halo. However, parametric strong lensing models for A1689 need at least a second matter clump at northwest together with the main halo to be effective. We did not try to model separately the halo of the brightest cluster galaxy (BCG) for two main reasons. Firstly, we were interested in total matter distribution and we did not try to distinguish between baryons and dark matter. This require the modelling of the intra-cluster medium too \citep{ser+al09}. Secondly, in order to effectively differentiate the BCG contribution from the main dark matter halo, we need dynamical data of the inner velocity profile \citep{san+al08}. 

We could parametrize the inner cluster profile as a main NFW halo plus an additional north-west component, modelled as a pseudo-Jaffe profile  \citep{kee01a}. As for weak lensing, we explored the parameter space by running four Markov chains of 50000 samples each. Parameters were diagonalized after an early run. Results are summarised in Table~\ref{tab_fit_nfw_sl_wl}. The average residual in image fitting was $\ls 2.5\arcsec$ in line with previous modelings \citep[ see table~1]{coe+al10}. 

Strong lensing parametric models may be oversimplified. A fitting model with the accuracy of the HST astrometric resolution should resolve substructure approximately 25~kpc across within the central $\sim 400~\mathrm{kpc}$ diameter \citep{coe+al10}. Simple models cannot account for such mass complexities. A coarser positional error allowed us to perform the lensing analysis without adding too many parameters and, at the same time, avoiding the explored region in parameter space being overly confined \citep{san+al08}. We took the following approach. First, we found a global minimum of the likelihood function ($\chi^2_\mathrm{SL}\sim 3231$ for a positional uncertainty of $\Delta \theta = 0.4\arcsec$). Since all images were provided with the same error, the chosen uncertainty does not affect the position of the minimum. Then, we rescaled the positional error such that the reduced $\chi^2_\mathrm{SL}=1$. Such rescaled error was then used in the Markov chain analysis.

Comparing weak to strong lensing results can be tricky. Strong lensing focuses on the cluster core and extrapolation well beyond the observational radius covered by multiple images is needed to infer global properties of the halo. On the other side, weak lensing observations provide a good coverage of the virial radius but are not accurate enough in the central regions to infer the inner slope of the density profile. Assuming the same parametric model for both strong and weak lensing assures us of interpreting the data coherently in the same framework, but we have still to check for consistency. Projected parameters inferred from either weak or strong lensing analysis of A1689 turn out to be in good agreement, see Table~\ref{tab_fit_nfw_sl_wl} and Figs.~\ref{fig_histo_ks_sl_wl_redsex},~\ref{fig_ks_rsP_sl_wl_redsex}. Marginalised posterior distributions for the halo strength, projected radius and ellipticity are plotted in Figs.~\ref{fig_histo_ks_sl_wl_redsex}. Even if the strong lensing analysis prefers smaller (larger) values for $k_\mathrm{s}$ ($r_\mathrm{sP}$, respectively), distributions are still compatible, as also enlightened by comparing the marginalised distribution $p(k_s, r_{sP})$, see Fig.~\ref{fig_ks_rsP_sl_wl_redsex}. The SL constraint on $k_{s}$ is much tighter than the WL one, whereas dispersions in the probability density functions (PDFs) for the projected radius are similar. On the other hand, the weak lensing constraint on ellipticity is not very strong, since only very large values of $\epsilon$ are ruled out. Being the WL PDF for ellipticity quite flat, the good agreement with the SL result is not very significant.

\section{Lensing inversion}
\label{sec_inve}

\begin{table*}
\begin{tabular}{@{}ccc | p{.3cm}p{.3cm}c | p{.3cm}p{.3cm}c|p{.3cm}p{.3cm}c|p{.3cm}p{.3cm}c|p{.3cm}p{.3cm}c|p{.3cm}p{.3cm}c}
        \hline
        \noalign{\smallskip}
	\multicolumn{3}{c}{Priors} & \multicolumn{3}{c}{$M_{200}$}	& \multicolumn{3}{c}{$c_{200}$}	& \multicolumn{3}{c}{$q_1$}	& \multicolumn{3}{c}{$q_2$}	& \multicolumn{3}{c}{$\cos\theta$}	& \multicolumn{3}{c}{$\cos\phi$}	 \\
        \noalign{\smallskip}
        $q_i$	&angles	&$c(M)$	&$\mu$	&$\sigma$	&$\gamma_1$	&$\mu$	&$\sigma$	&$\gamma_1$	&$\mu$	&$\sigma$	&$\gamma_1$	&$\mu$	&$\sigma$	&$\gamma_1$	&$\mu$	&$\sigma$	&$\gamma_1$	&$\mu$	&$\sigma$	&$\gamma_1$ \\
        \hline
         \multicolumn{18}{c}{Weak lensing}	 \\
         \noalign{\smallskip}
	flat	&random	&$\times$	&$14.8$	&$5.1$	&$0.4$	&$12.7$	&$4.0$	&$1.0$	&$0.68$	&$0.25$	&$-.7$	&$0.88$	&$0.11$	&$-1.4$	&$0.49$	&$0.30$	&$0.0$	&$0.57$	&$0.32$	&$-.1$ \\
	flat	&bias	&$\times$	&$13.9$	&$4.6$	&$0.3$	&$11.3$	&$3.0$	&$1.5$	&$0.66$	&$0.22$	&$-.7$	&$0.83$	&$0.17$	&$-.9$	&$0.91$	&$0.07$	&$-.9$	&$0.56$	&$0.33$	&$-.1$ \\
	flat	&random	&$\checkmark$		&$5.9$	&$4.0$	&$1.0$	&$5.5$	&$1.7$	&$0.6$	&$0.66$	&$0.19$	&$-.6$	&$0.85$	&$0.14$	&$-1.3$	&$0.61$	&$0.30$	&$-.4$	&$0.71$	&$0.29$	&$-.7$ \\
	$N$-body	&random	&$\times$	&$14.6$	&$4.8$	&$0.2$	&$12.1$	&$3.5$	&$0.6$	&$0.49$	&$0.08$	&$0.1$	&$0.69$	&$0.12$	&$-.1$	&$0.60$	&$0.30$	&$-.4$	&$0.45$	&$0.30$	&$0.3$ \\
        $N$-body	&bias	&$\times$	&$13.0$	&$4.2$	&$0.4$	&$10.2$	&$2.8$	&$0.6$	&$0.46$	&$0.09$	&$0.0$	&$0.59$	&$0.13$	&$0.2$	&$0.92$	&$0.06$	&$-1.1$	&$0.58$	&$0.32$	&$-.2$ \\  
         $N$-body	&bias	&$\checkmark$	&$7.2$	&$4.0$	&$0.7$	&$5.3$	&$1.8$	&$0.7$	&$0.44$	&$0.10$	&$-.1$	&$0.56$	&$0.15$	&$0.0$	&$0.94$	&$0.05$	&$-1.5$	&$0.65$	&$0.32$	&$-.4$ \\
\hline
        \multicolumn{18}{c}{Strong lensing}	 \\
        \noalign{\smallskip}
        flat	&random	&$\times$	&$21.4$	&$5.3$	&$0.8$	&$7.9$	&$2.2$	&$1.7$	&$0.46$	&$0.22$	&$-.4$	&$0.77$	&$0.14$	&$-.4$	&$0.49$	&$0.31$	&$0.0$	&$0.52$	&$0.31$	&$0.0$ \\
	flat	&bias	&$\times$	&$18.5$	&$3.9$	&$0.5$	&$6.5$	&$1.0$	&$2.5$	&$0.49$	&$0.17$	&$-.6$	&$0.75$	&$0.19$	&$-.4$	&$0.90$	&$0.07$	&$-.8$	&$0.64$	&$0.30$	&$-.3$ \\
	flat	&random	&$\checkmark$		&$21.0$	&$3.7$	&$0.0$	&$6.3$	&$0.9$	&$0.8$	&$0.59$	&$0.12$	&$-.8$	&$0.79$	&$0.16$	&$-.7$	&$0.65$	&$0.30$	&$-.5$	&$0.81$	&$0.23$	&$-1.1$ \\
$N$-body	&random	&$\times$	&$18.8$	&$4.0$	&$0.2$	&$7.2$	&$1.3$	&$0.5$	&$0.48$	&$0.08$	&$0.0$	&$0.66$	&$0.12$	&$-.1$	&$0.67$	&$0.28$	&$-.7$	&$0.61$	&$0.28$	&$-.3$ \\
    $N$-body	&bias	&$\times$	&$17.2$	&$3.4$	&$0.0$	&$6.1$	&$0.9$	&$0.3$	&$0.43$	&$0.08$	&$-.2$	&$0.58$	&$0.14$	&$0.1$	&$0.91$	&$0.06$	&$-.7$	&$0.72$	&$0.28$	&$-.8$ \\
         $N$-body	&bias	&$\checkmark$	&$18.2$	&$3.0$	&$0.0$	&$5.4$	&$0.8$	&$0.4$	&$0.42$	&$0.09$	&$0.1$	&$0.56$	&$0.14$	&$0.3$	&$0.95$	&$0.05$	&$-1.3$	&$0.75$	&$0.29$	&$-.9$ \\
        \hline
        \multicolumn{18}{c}{Weak and Strong lensing}	 \\
        \noalign{\smallskip}
	flat	&random	&$\times$	&$15.3$	&$3.3$	&$0.8$	&$9.3$	&$2.0$	&$1.8$	&$0.48$	&$0.20$	&$-.5$	&$0.77$	&$0.14$	&$-.3$	&$0.47$	&$0.28$	&$0.1$	&$0.60$	&$0.30$	&$-.3$ \\
	flat	&bias	&$\times$	&$14.1$	&$2.4$	&$0.8$	&$7.7$	&$1.1$	&$2.2$	&$0.45$	&$0.18$	&$-.5$	&$0.77$	&$0.18$	&$-.6$	&$0.90$	&$0.07$	&$-.7$	&$0.58$	&$0.31$	&$-.2$ \\
	flat	&random	&$\checkmark$		&$14.4$	&$2.2$	&$0.1$	&$7.6$	&$1.0$	&$-.3$	&$0.55$	&$0.14$	&$-1.0$	&$0.76$	&$0.20$	&$-.9$	&$0.81$	&$0.26$	&$-1.1$	&$0.78$	&$0.31$	&$-.8$ \\
	$N$-body	&random	&$\times$	&$14.9$	&$2.2$	&$0.1$	&$9.0$	&$1.2$	&$-.1$	&$0.48$	&$0.08$	&$0.1$	&$0.66$	&$0.10$	&$0.1$	&$0.52$	&$0.26$	&$-.2$	&$0.48$	&$0.29$	&$0.1$ \\
	$N$-body	&bias	&$\times$	&$13.4$	&$2.0$	&$0.0$	&$7.3$	&$0.8$	&$0.1$	&$0.45$	&$0.08$	&$-.1$	&$0.62$	&$0.14$	&$0.0$	&$0.90$	&$0.07$	&$-.8$	&$0.72$	&$0.27$	&$-.8$ \\
	$N$-body	&bias	&$\checkmark$	&$13.3$	&$1.8$	&$0.1$	&$6.6$	&$0.8$	&$0.0$	&$0.42$	&$0.09$	&$-.3$	&$0.57$	&$0.15$	&$0.0$	&$0.94$	&$0.05$	&$-1.2$	&$0.84$	&$0.20$	&$-1.4$ \\
  \hline
\end{tabular}
\caption{Intrinsic parameters inferred through lensing inversion assuming different priors. $\mu$ and $\sigma$ are the biweight estimators for location and scale of the PDF \citep{bee+al90}; $\gamma_1$ is the skewness. Masses are in units of $10^{14}M_\odot$.}
\label{tab_pdf_par}
\end{table*}

Deprojecting a surface density map to infer the intrinsic 3D shape is an under-constrained astronomical problem. Let us consider the ellipsoidal NFW halo. The 3D halo feature six parameters: the density profile is described by two parameters ($M_{200}$ and $c_{200}$), the shape by two axial ratios ($q_1$ and $q_2$), the orientation by two angles, $\theta$ and $\phi$\footnote{We are neglecting the third angle $\psi$ that fixes the direction in the plane of the sky and is related to the orientation angle $\theta_\epsilon$}. On the other hand, from lensing observations, we only get three constraints relating the intrinsic parameters: the lensing strength $k_\mathrm{s}$, the projected radius $r_\mathrm{sP}$, and the ellipticity $\epsilon$. Even combining multi-wavelength observations, from X-ray to optical to radio band, one can only constrain the elongation of the cluster along the line of sight \citep{ser07}.

One way to address such shortcomings is to exploit some prior information \citep{ogu+al05,cor+al09}. The Bayes theorem states that
\beq
\label{baye}
p(\bfP | \bfd) \propto {\cal L}( \bfP|\bfd) p(\bfP),
\eeq
where $p(\bfP | \bfd)$ is the posterior probability of the parameters $\bfP$ given the data $\bfd$, ${\cal L}( \bfP|\bfd)$ is the likelihood of the data given the model parameters and $p(\bfP)$ is the prior probability distribution for the model parameters. The statistical approach described above is the same of \citet{ogu+al05} and \citet{cor+al09}, but our method differs in some important points as far as the strong lensing analysis is concerned. \citet{cor+al09} only analysed weak lensing data, whereas \citet{ogu+al05} used a circularly averaged profile of the convergence map of the inner regions to put strong lensing constraints. In the present paper, we have performed a parametric analysis of the multiple image systems so that we can employ a strong lensing likelihood that is fully triaxial. 

The likelihood for the intrinsic parameters is the same as in Sec.~\ref{sec_like_sl} and \ref{sec_like_wl} for strong or weak lensing analyses, respectively. In fact, the three projected parameters can be expressed as functions of the intrinsic ones, see Sec.~\ref{sec_nfw}.

We considered several priors, see Sec.~\ref{sec_theo}. For the axial ratios $q_1$ and $q_2$, we considered either the $N$-body predictions in Eqs.~(\ref{nbod3}--\ref{nbod4}) or a flat distribution, see Eqs.~(\ref{flat1},~\ref{flat2}). We always put a lower bound $q_1 \ge 0.05$. For the alignment angle $\theta$, we considered either the biased distribution for $p(\theta)$ in Eq.~(\ref{nbod5}) or a random distribution, Eq.~(\ref{flat3}). For the azimuth angle $\varphi$ we always used a random flat distribution, $p(\varphi) = const$.. For the mass, we always used a flat prior $p(M_{200}) = const.$, whereas the a priori PDF for the concentration was flat in the range $0 < c_{200} \le 30$ and null otherwise. The $c(M)$ relation, see Eq.~(\ref{nbod2}), was either enforced or neglected. Posterior PDFs were computed by running four chains for each of the three cases (only WL, only SL, combined WL and SL). We computed at least 20000 samples per chain and eventually added ten thousands more until the convergence criterium was satisfied.

\subsection{Weak lensing}

\begin{figure*}
\includegraphics[width=18cm]{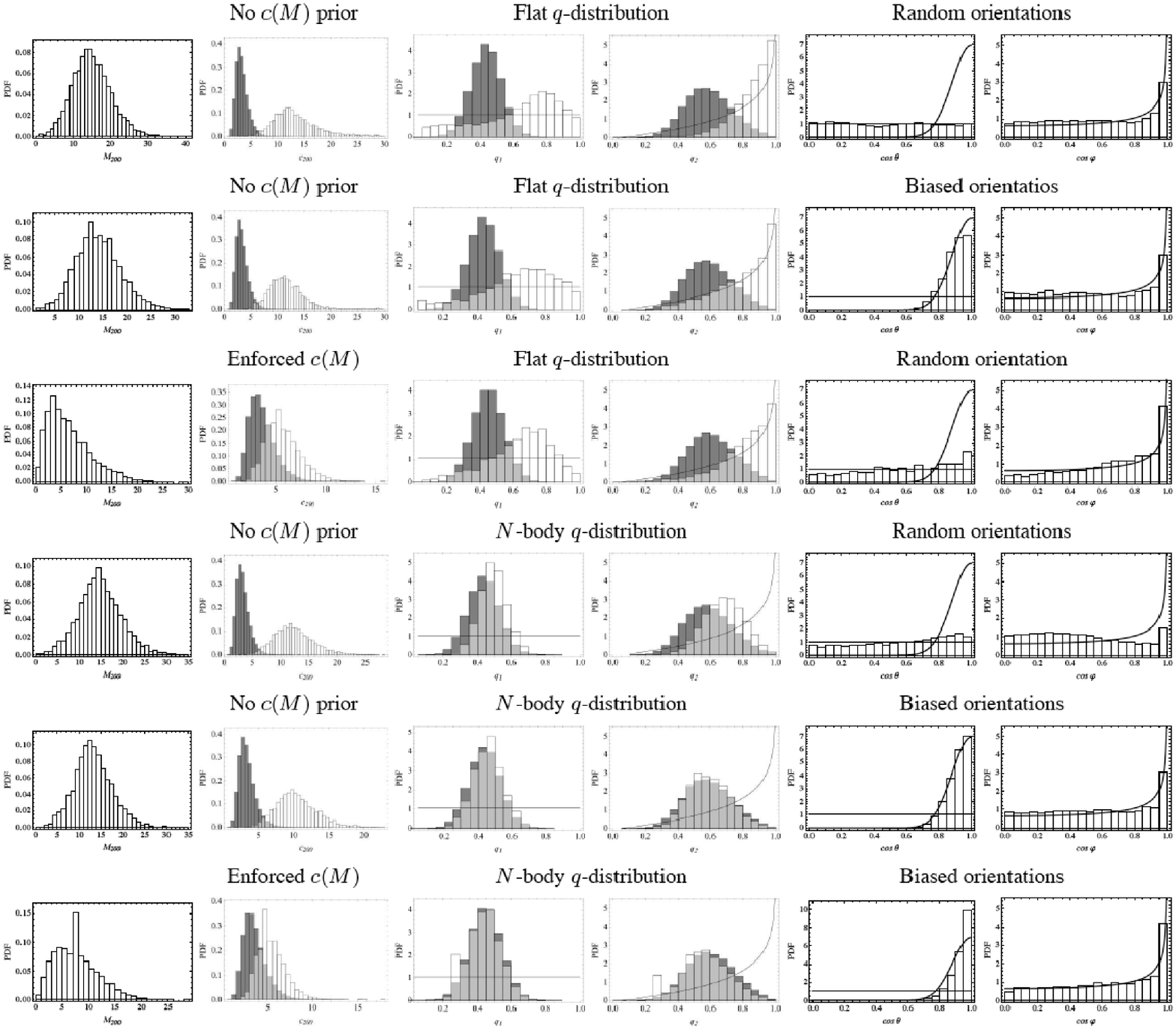} 
\caption{PDFs for the intrinsic halo parameters as derived from weak lensing inversion. Panels from the left to the right are for $M_{200}$, $c_{200}$, $q_1$, $q_2$, $\cos \theta$ and $\cos \varphi$, respectively. The white binned PDF is the posterior PDF; the dark grey histogram is the distribution expected according to predictions from $N$-body simulations given the mass distribution in the left panel; intermediate grey bins are shared by the two plotted distributions. Thin lines in the $q$-panels denote a flat $q$-distribution. Thin lines in the $\theta$ and $\phi$-angle panel denote a random orientation. The thick line in the $\cos \theta$-panel represents a biased orientation. Priors are stated above each set of PDFs. From the top to the bottom row, we plot the weak lensing inversion results under the prior hypotheses of: flat $q$-distribution and random orientation angles; flat $q$-distribution and biased orientation angles; flat $q$-distribution, random orientation angles and fulfilled $c(M)$ relation; $N$-body $q$-distribution and random orientation angles; $N$-body $q$-distribution and biased orientation angles; $N$-body $q$-distribution, biased orientation angles and enforced $c(M)$ relation, respectively. Masses are in units of $10^{14}M_\odot$.}
\label{fig_pdf_WL}
\end{figure*}

One of the main sources of bias in the lensing-based $c(M)$ measurements is that an average triaxial halo with the major axis oriented along the line of sight appears to be over-concentrated if analysed under the assumption of spherical symmetry \citep{ogu+al05,co+ki07}. Allowing for triaxial shapes and casual orientation, the inferred distribution of either mass or concentration will come out much shallower than under the sharp hypothesis of spherical symmetry \citep{cor+al09}. 

We retrieve these trends. Results are summarised in Table~\ref{tab_pdf_par}. In Fig.~\ref{fig_pdf_WL} we plot as binned histograms the marginalised posterior probability density functions for the intrinsic halo parameters, computed exploiting different a priori hypotheses. For each parameter, we compared the inferred PDF with the theoretical prediction from $N$-body simulations, see Sec.~\ref{sec_theo}. Predictions for concentration and axial ratios were computed by convolution of $N$-body predictions in Eqs.~(\ref{nbod1},~\ref{nbod2}~\ref{nbod3}, and~\ref{nbod4}) with the inferred posterior PDF for the mass, plotted in the leftmost row. The expected and inferred $c_{200}$ distributions have different meanings. The theoretical one is the concentration function of a population of clusters with a mass function as in the leftmost panel; the inferred one refers to the probability density for the concentration of a single cluster (A1689) with a given mass (to be determined). The biased distribution for the orientation angle $\theta$ was plotted as a continuos thick line (the same representation used for the random distribution of $\phi$). The flat distributions for $q_1$ and $q_2$ and the random one for $\theta$ were plotted as thin lines for comparison.

The assumptions of either mildly triaxial halos, as suggested by simulations, or biased orientations have similar effects on the concentrations. The tail of the distribution at large values gets thinner whereas the maximum shifts towards lower values. Only by enforcing the $c(M)$ relation, the $c_{200}$ distribution becomes very similar to the theoretical one, even if over-concentrated values are still slightly preferred.

Since WL is not effective in determining the projected ellipticity, constraints on the intrinsic axial ratios are not sharp and are dominated by the a priori hypothesis employed. However, a very elongated shape, $q_1\ls 0.2$, is strongly disfavoured with respect to mildly triaxial configurations with $q_1\gs 0.3$. Assuming flat distributions, the mode of the distribution for $q_1$ is in the range 0.7 -- 0.8, whereas using priors from $N$-body simulations, the inferred distributions become very similar to the theoretical ones, with peak values $q_1\sim0.4-0.5$. 

The results on the second axial ratio $q_2$ and the angle $\phi$ are not very significant since they are dominated by the prior. Results on the inclination angle of the major axis, $\theta$, are more interesting. When assuming random orientations, aligned configurations ($\cos \theta \ls 1$) are not significantly preferred. The peak of the PDF moves towards smaller values of $\theta$ if either mildly triaxial shapes are enforced or the $c(M)$ relation is assumed to be valid.

\subsection{Strong lensing}

\begin{figure*}
\includegraphics[width=18cm]{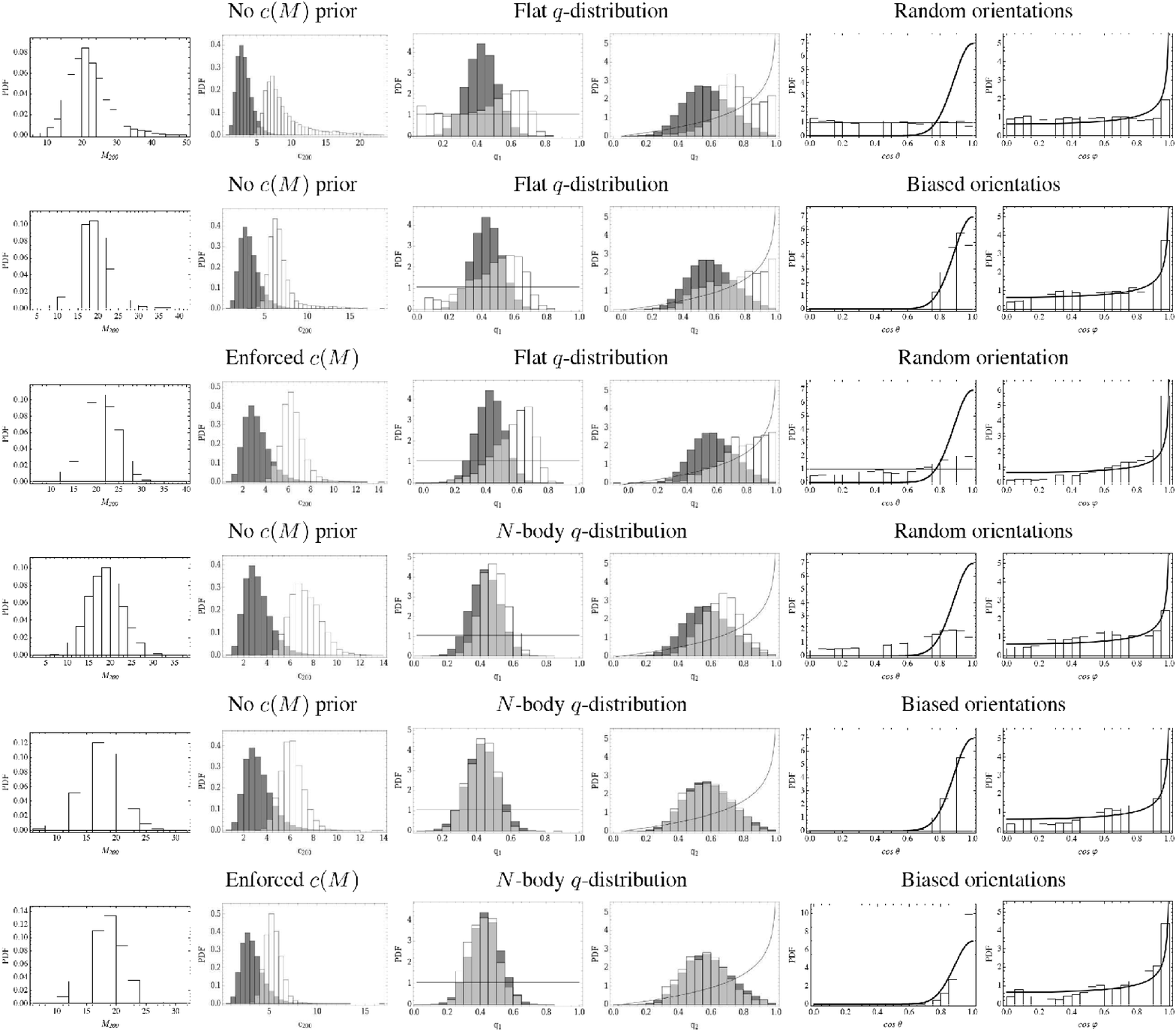} 
\caption{PDFs for the intrinsic halo parameters as derived from strong lensing inversion. Notation and style follow Fig.~\ref{fig_pdf_WL}.
}
\label{fig_pdf_SL}
\end{figure*}

Strong lensing inversion was performed by deprojecting the central NFW halo found in the analysis of the multiple image systems. The likelihood was computed as ${\cal L}_\mathrm{SL} \propto \exp \left\{-\chi^2_\mathrm{SL}/2 \right\}$. Marginalised posterior probability density functions for intrinsic parameters are plotted in Fig.~\ref{fig_pdf_SL}; corresponding momenta are listed in Table~\ref{tab_pdf_par}. As far as the effects of priors are concerned, we retrieve the same trends found in the weak lensing analysis. The cluster appears to be over-concentrated but still compatible with predictions. Assuming a priori either mildly triaxial axial ratios or biased orientations, the disagreement gets smaller. 

Strong lensing gives the more effective constraint on the projected ellipticity. Triaxiality increases the lensing strength, so halo models inferred from SL do not have to be extremely massive or concentrated. PDFs for mass and concentration obtained with strong lensing inversion are compatible but more peaked and less shallow than distributions from weak lensing. The concentration PDFs peak at lesser values. Central momenta for $c_{200}$ shift from the range $13-6$ to values between $6-5$; PDF dispersions obtained with SL inversion are nearly one half of those in the WL case ($1-2$ compared to $2-4$). There are still long tails at large values, but they do not extend behind $c_{200}\gs 15$. Strong lensing also prefers larger mass values, $\Delta M_{200} \sim (4-6)\times 10^{14}M_\odot$ for the central momenta with respect to WL. The shift in mass is even more pronounced when the prior on the $c(M)$ relation is enforced.

The prior on the $c(M)$ relation affects the weak or the strong lensing inversion to a different degree. Since weak lensing PDFs span large ranges in both mass and concentration, the $c(M)$ prior can dominate the posterior result by selecting a region which is always compatible with the likelihood. Then, the shapes of the PDFs for mass and concentration with or without the prior on the $c(M)$ relation are very different. On the other hand, the concentration distribution from SL is already peaked and in good agreement with theoretical predictions. Then, any $c(M)$ prior alleviates the over-concentration problem for the strong lensing analysis, favouring lesser values of $c_{200}$ and smaller dispersions, but does not dramatically change the PDFs.

Differently from any shear analysis in the outskirts, strong lensing can precisely estimate the projected ellipticity in the inner regions. Whatever the priors on the axial ratios distributions, spherical shapes are quite disregarded. Even assuming a priori flat distributions, values of $q_1 \gs 0.8$ are now excluded and final distributions are quite peaked favouring triaxial shapes. Using strong instead of weak lensing, the central momenta of the PDF for $q_1$ and $q_2$ shift by $\sim0.2$ and $\sim0.1$, respectively. As before, when we adopt $N$-body like prescriptions, posterior PDFs mimic the prior.

\subsection{Weak plus strong lensing}

\begin{figure*}
\includegraphics[width=18cm]{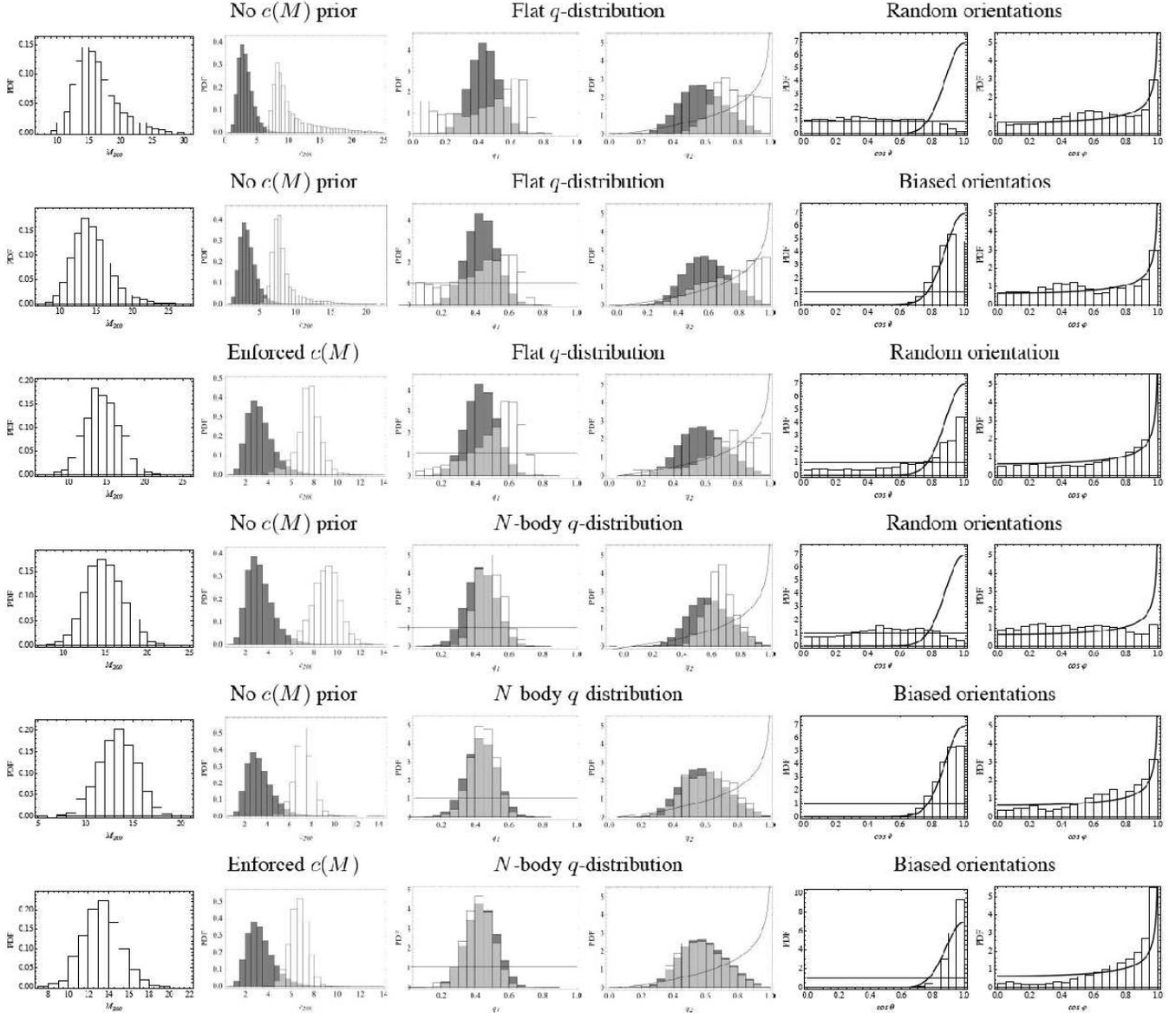} 
\caption{PDFs for the intrinsic halo parameters as derived from combined weak and strong lensing inversion. Notation and style follow Fig.~\ref{fig_pdf_WL}.
}
\label{fig_pdf_All}
\end{figure*}

In order to perform a weak+strong lensing inversion, we considered a combined likelihood ${\cal L}_\mathrm{All} \propto {\cal L}_\mathrm{WL} \times {\cal L}_\mathrm{SL}$. Since posterior distributions from either strong or weak lensing analyses are compatible, the combined investigation provides a meaningful compromise. 

Concentration values are a bit larger but still compatible with theoretical predictions. With respect to the inversion using only strong lensing, lesser values of the mass are preferred, $-\Delta M_{200} \sim (4-6) \times 10^{14}M_{\odot}$, with smaller dispersions. On the other hand, concentrations are slightly higher, $\Delta c_{200} \sim 1-1.5$, but with similar dispersions and shapes. Since strong lensing gives a more effective constraint on the projected ellipticity, the posterior distributions for the axial ratios are quite similar to the strong lensing results.

\section{Results}
\label{sec_resu}

Several features of the inferred distributions of intrinsic parameters are more easily understood by considering multi-dimensional posterior PDFs. Some trends are quite clear.

\subsection{$c(M)$ relation}

\begin{figure*}
\begin{center}
$
\begin{tabular}{c@{\hspace{.2cm}}c@{\hspace{.2cm}}c}
   Weak lensing		&Strong lensing	&Weak and Strong lensing \\
   \noalign{\smallskip}
\includegraphics[width=5cm]{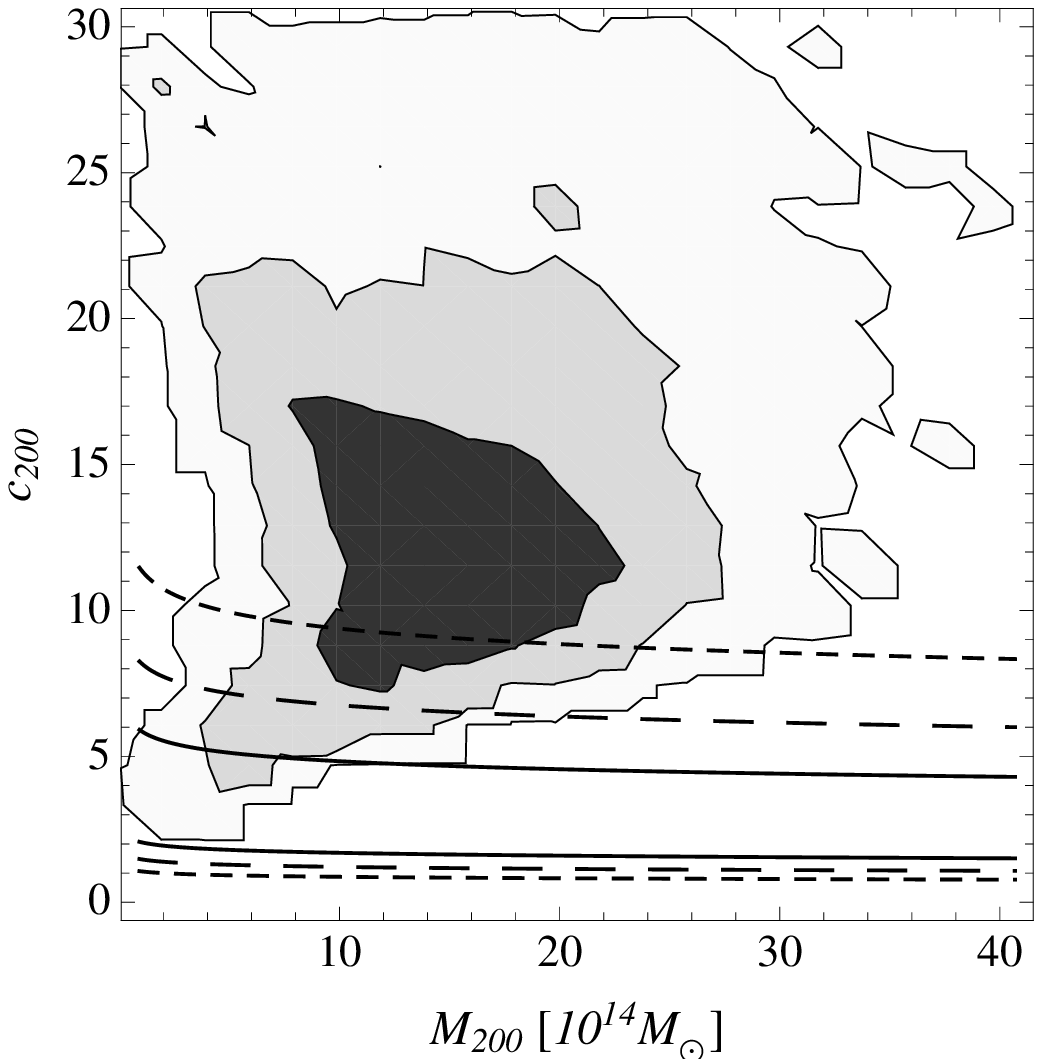} &
\includegraphics[width=5cm]{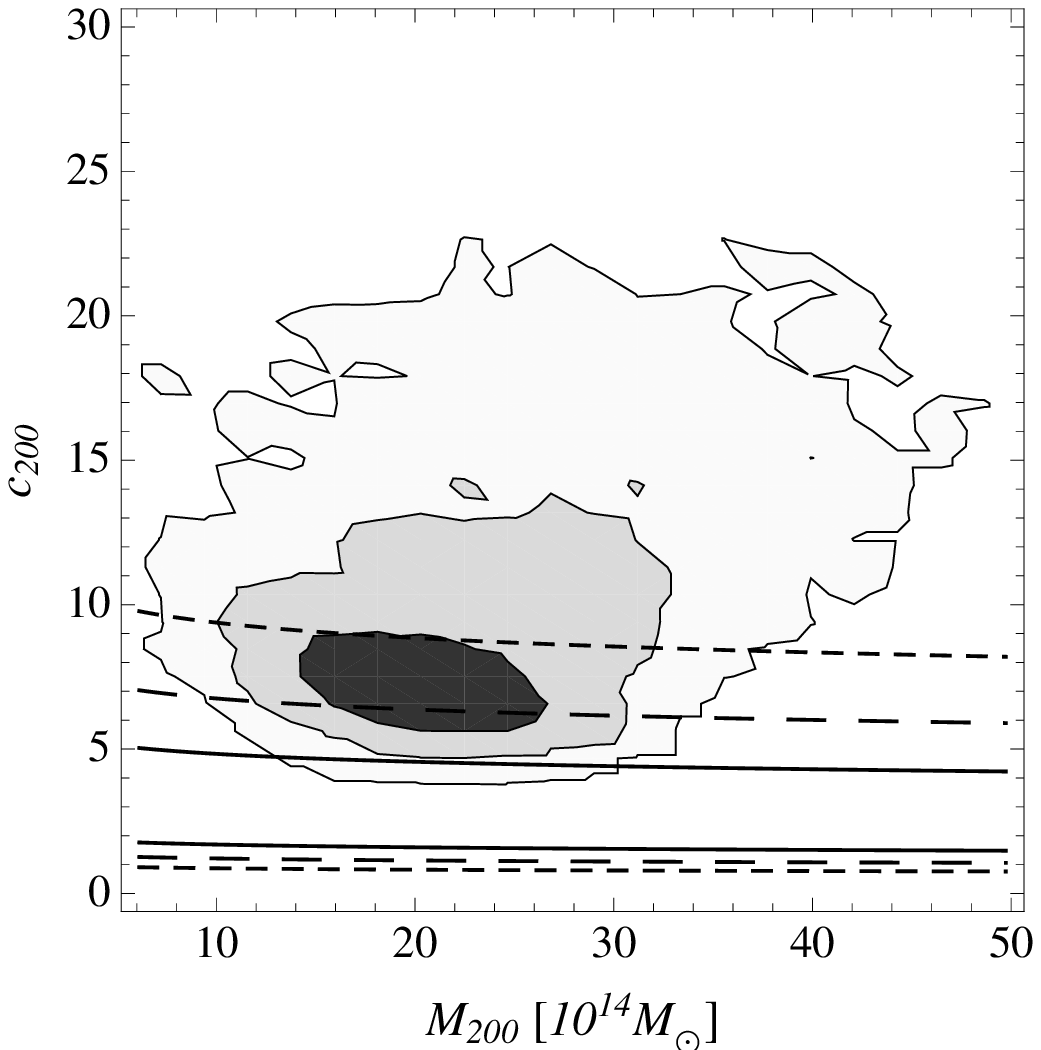} &
\includegraphics[width=5cm]{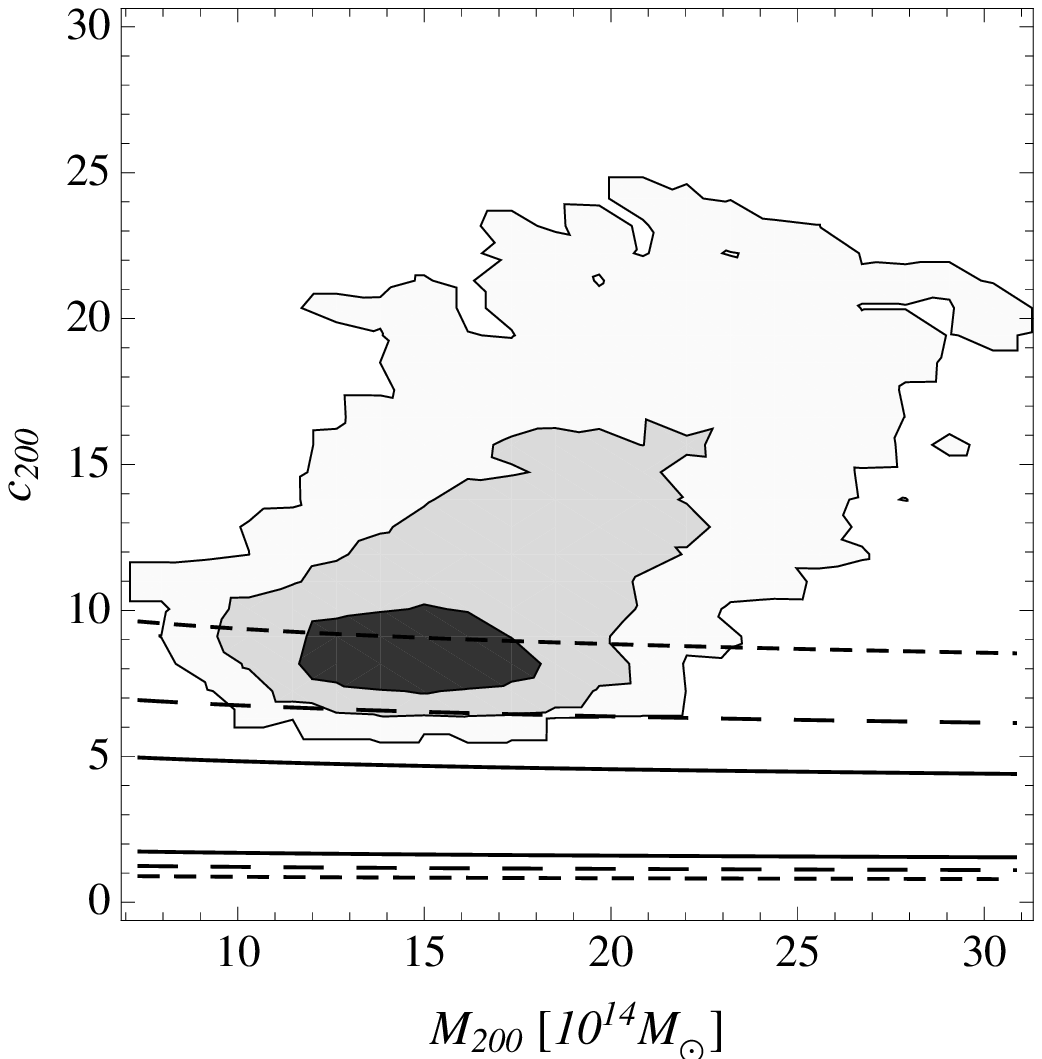} \\
\end{tabular}
$
\end{center}
\caption{Contour plot of the marginalised PDF for $M_{200}$ and $c_{200}$ as derived under the prior assumptions of flat $q$-distribution and random orientation angles. 
Panels from left to right are for the weak, strong and combined lensing inversion, respectively. Contours are plotted at fraction values $\exp (-2.3/2)$, $\exp(-6.17/2)$, and $\exp(-11.8/2)$ of the maximum, which denote confidence limit regions of 1, 2 and $3\sigma$ in a maximum likelihood investigation, respectively. The full, long-dashed and dashed lines enclose the 1, 2 and $3\sigma$ regions for the predicted conditional probability $c(M)$, respectively.
}
\label{fig_M200_c200_flat_random_noprior}
\end{figure*}

\begin{figure*}
\begin{center}
$
\begin{tabular}{c@{\hspace{.2cm}}c@{\hspace{.2cm}}c}
   Weak lensing		&Strong lensing	&Weak and Strong lensing \\
   \noalign{\smallskip}
\includegraphics[width=5cm]{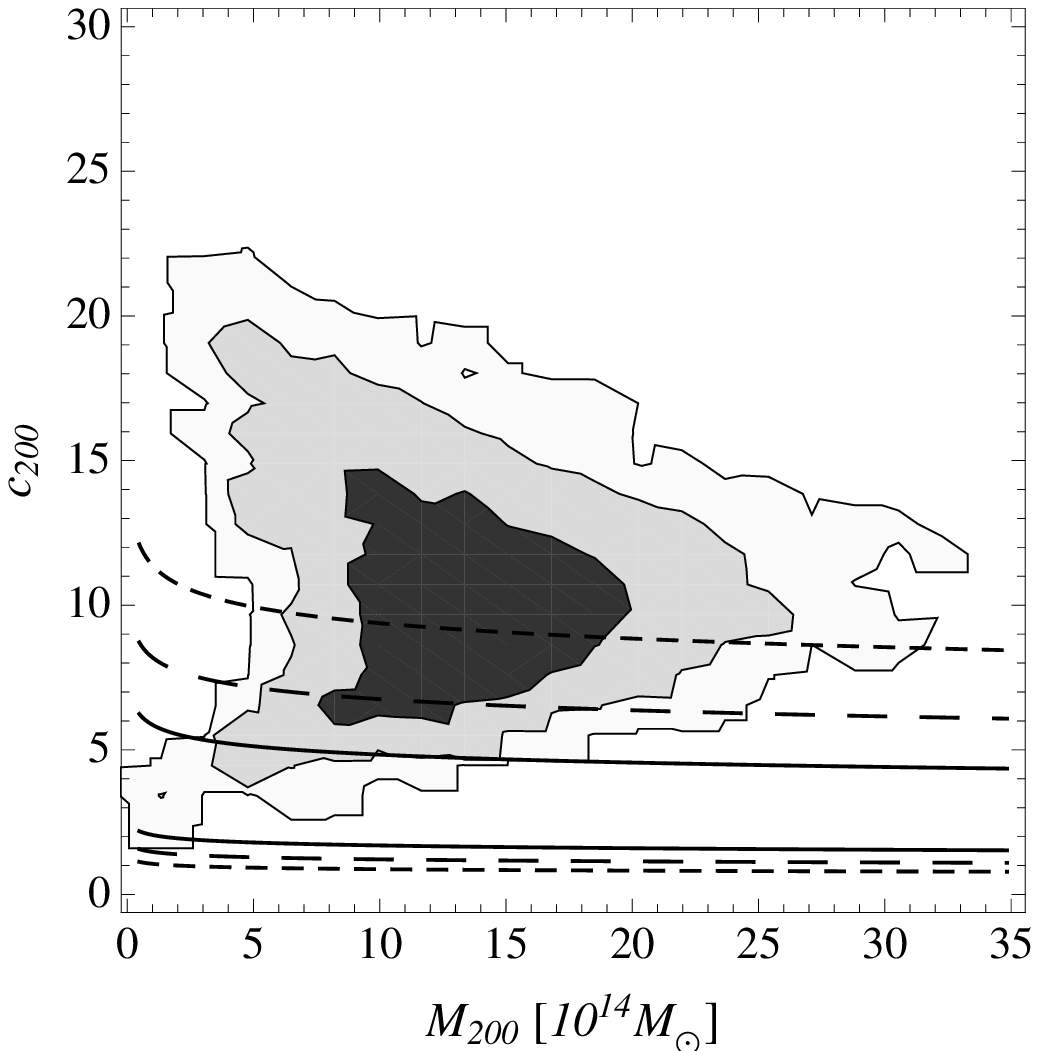} &
\includegraphics[width=5cm]{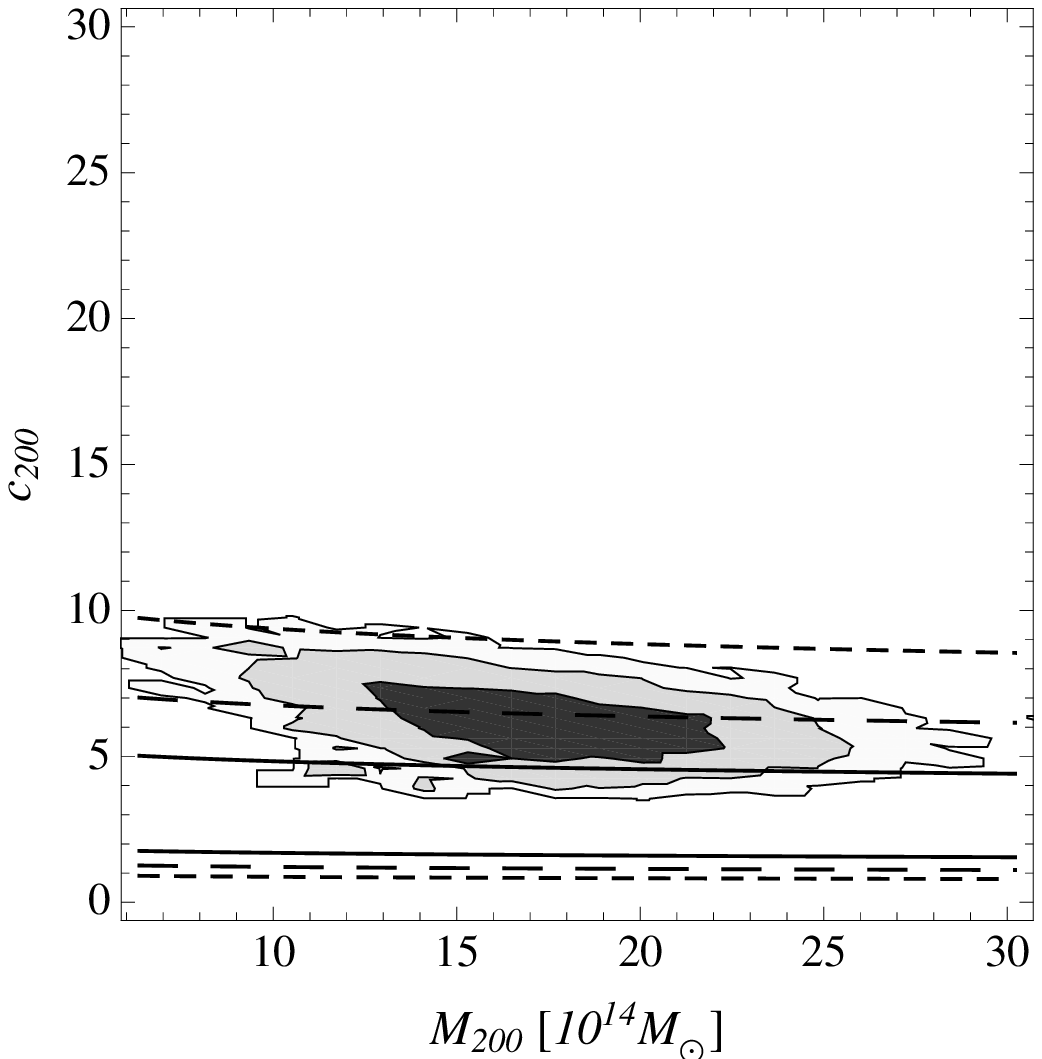} &
\includegraphics[width=5cm]{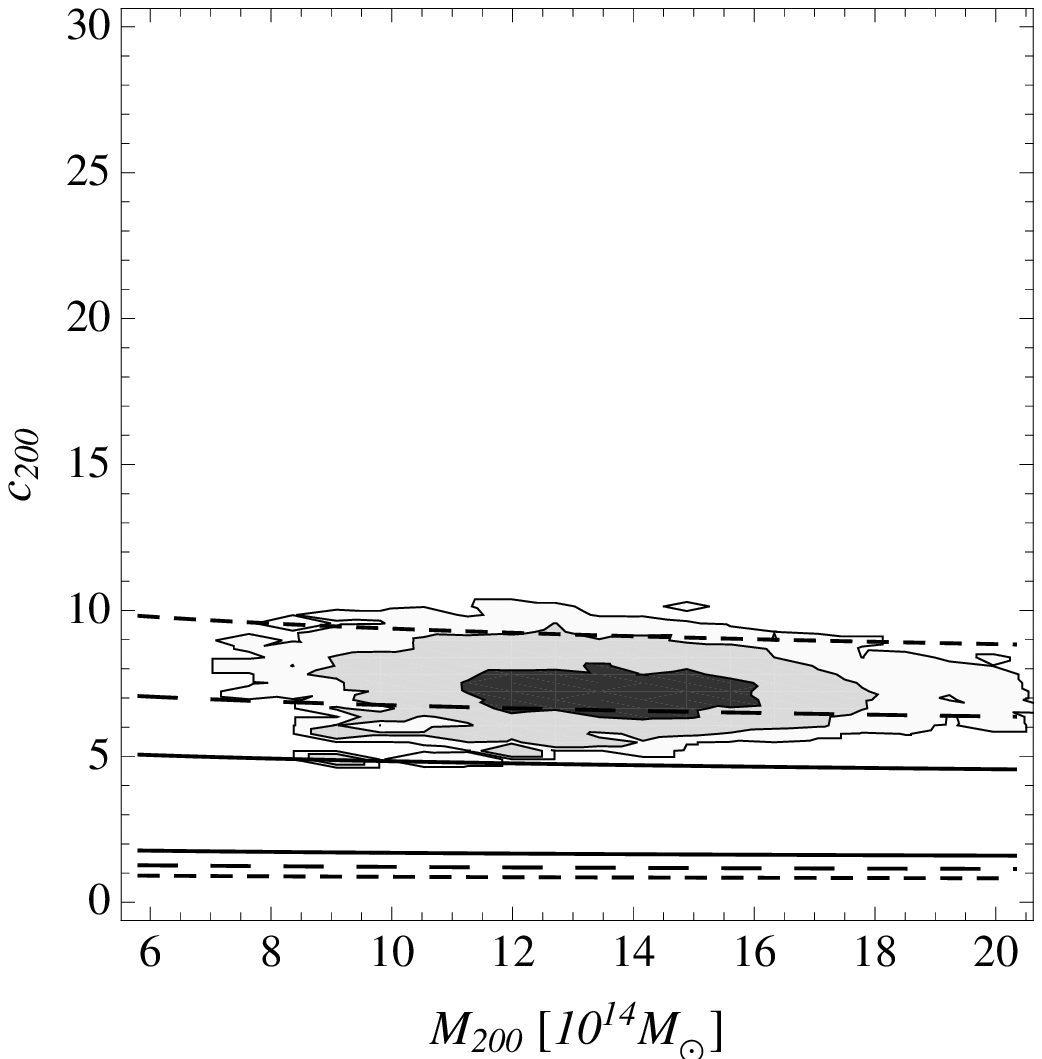} \\
\end{tabular}
$
\end{center}
\caption{The same as Fig.~\ref{fig_M200_c200_flat_random_noprior} when assuming an $N$-body like $q$-distribution and biased orientation angles.}
	\label{fig_M200_c200_nbody_bias_noprior}
	\end{figure*}

Whatever the assumptions on either orientation or shape, there is agreement between theoretical predictions and measured values of mass and concentration. This can be seen analysing the marginalised $p(M_{200},c_{200})$, see Figs.~\ref{fig_M200_c200_flat_random_noprior} and~\ref{fig_M200_c200_nbody_bias_noprior}, where we compared the computed posterior PDF with the theoretical conditional probability for the concentration given a halo mass. 

A1689 appears to be a quite typical massive cluster with a concentration in agreement with the tail at large values of the expected population of clusters of that given mass. The peak of $p(M_{200},c_{200})$ as inferred from strong (weak) lensing inversion is $\ls 1~(2)\sigma$ away from the maximum of the cluster concentration function, with the $1\sigma$ tail at smaller values compatible with the median expected value. This results is independent on the assumed prior. In fact, Figs.~\ref{fig_M200_c200_flat_random_noprior} and~\ref{fig_M200_c200_nbody_bias_noprior} consider two quite different scenarios: a priori flat distributions and random orientation versus axial ratios and inclination predicted by $N$-body simulations. 

The combined weak plus strong lensing analysis further constrains the preferable region in the $M_{200}$-$c_{200}$ plane. The tail at low concentrations is thinned, but there is still agreement. Independently of the priors, the inferred $c_{200}$ are only $\gs$ 1$\sigma$ away from the predicted median value. Priors from $N$-body simulations also help to put an upper bound on the concentrations. The $3\sigma$ region for the inferred $p(M_{200},c_{200})$ in the weak plus strong lensing is within $c_{200}\ls 11$.

As theoretical prediction we considered results from the analysis of a full sample of clusters \citep{duf+al08}. The $c(M)$ relation for a sample of only relaxed halos has a peak at larger concentrations and a smaller dispersion at a given mass. The inclusion of unrelaxed haloes adds a tail towards low concentrations. Even comparing the A1689 concentration and mass with results for relaxed clusters, we would have found agreement as well. Two opposite effects counterbalance if we consider the relaxed instead of the full sample. On one side, the distance between the peaks of either the inferred or the expected distribution gets smaller; on the other hand, the predicted tail al large values is smaller too, reducing the distribution overlapping.

\subsection{Shape}

\begin{figure*}
\begin{center}
$
\begin{tabular}{c@{\hspace{.2cm}}c@{\hspace{.2cm}}c}
   Weak lensing		&Strong lensing	&Weak and Strong lensing \\
   \noalign{\smallskip}
\includegraphics[width=5cm]{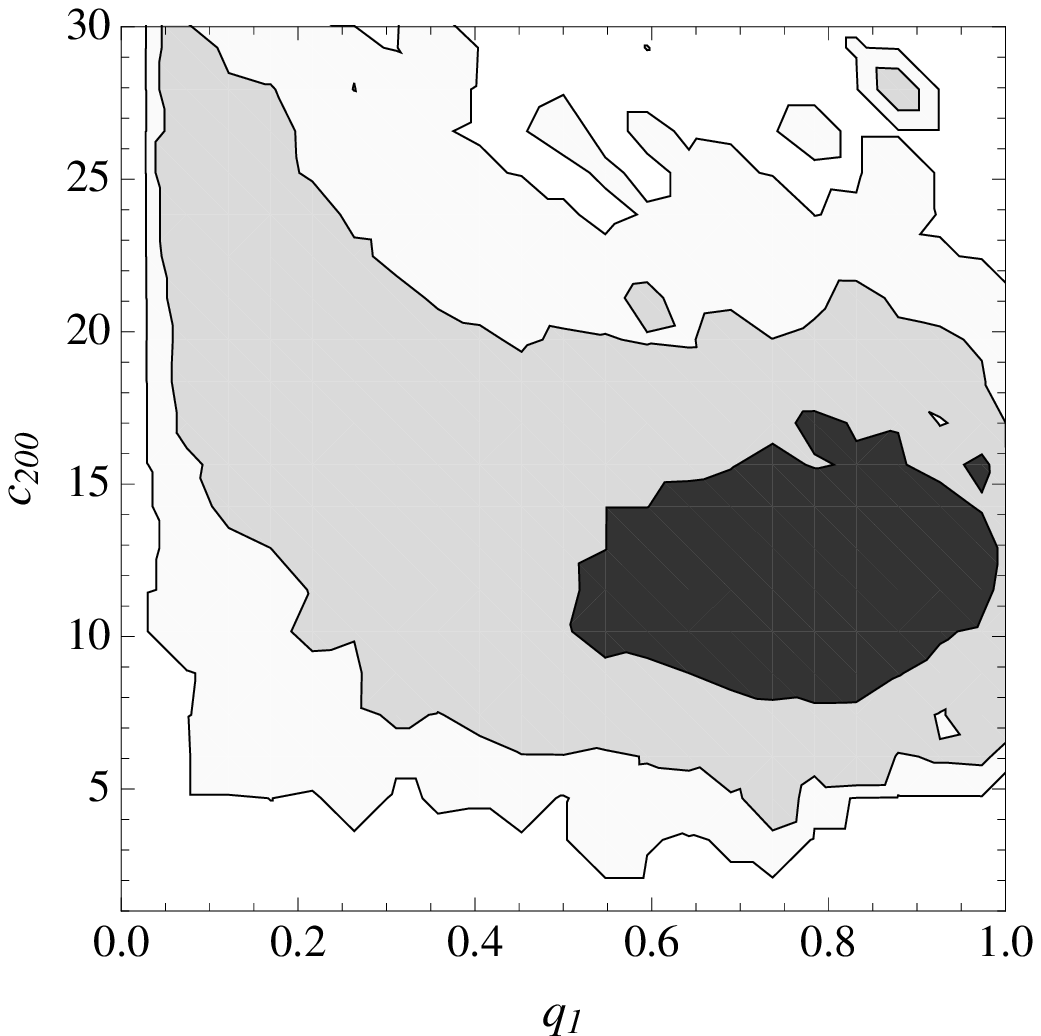} &
\includegraphics[width=5cm]{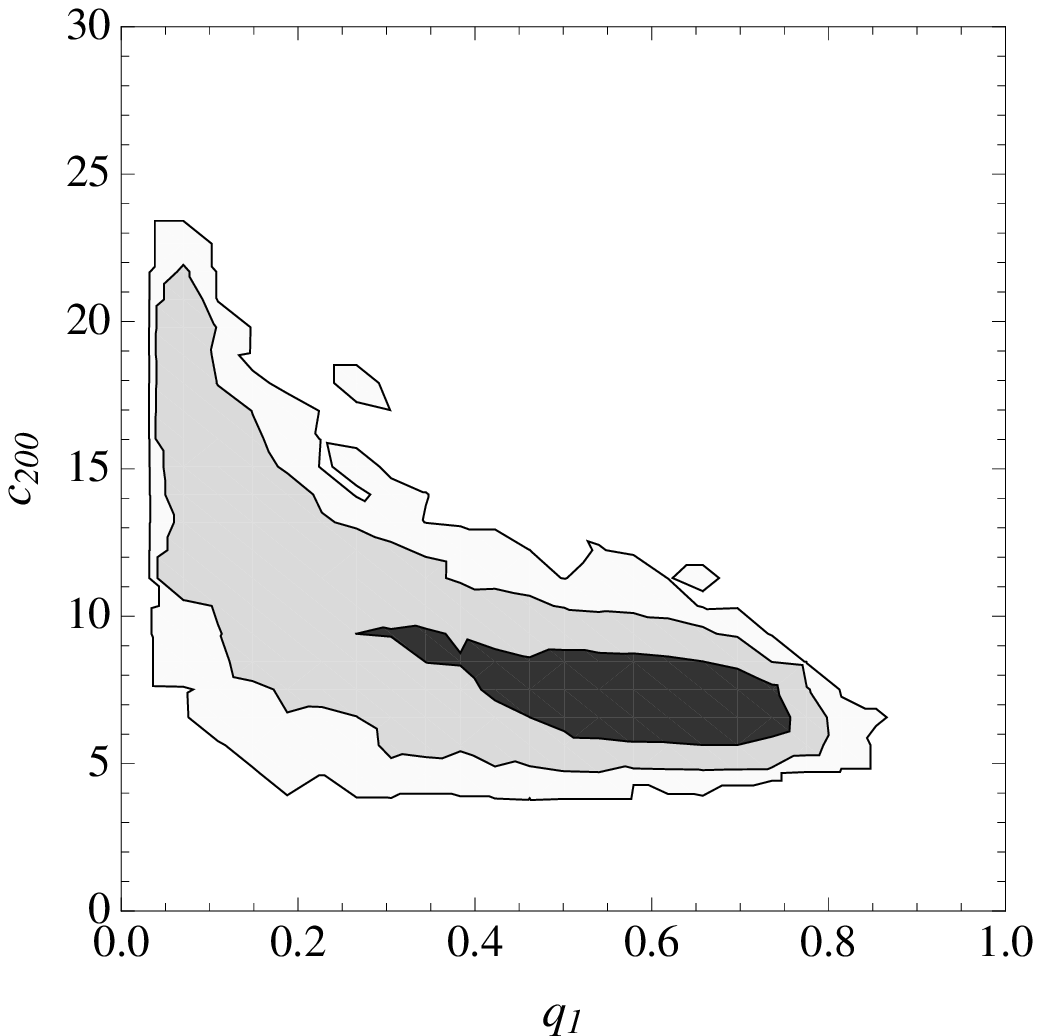} &
\includegraphics[width=5cm]{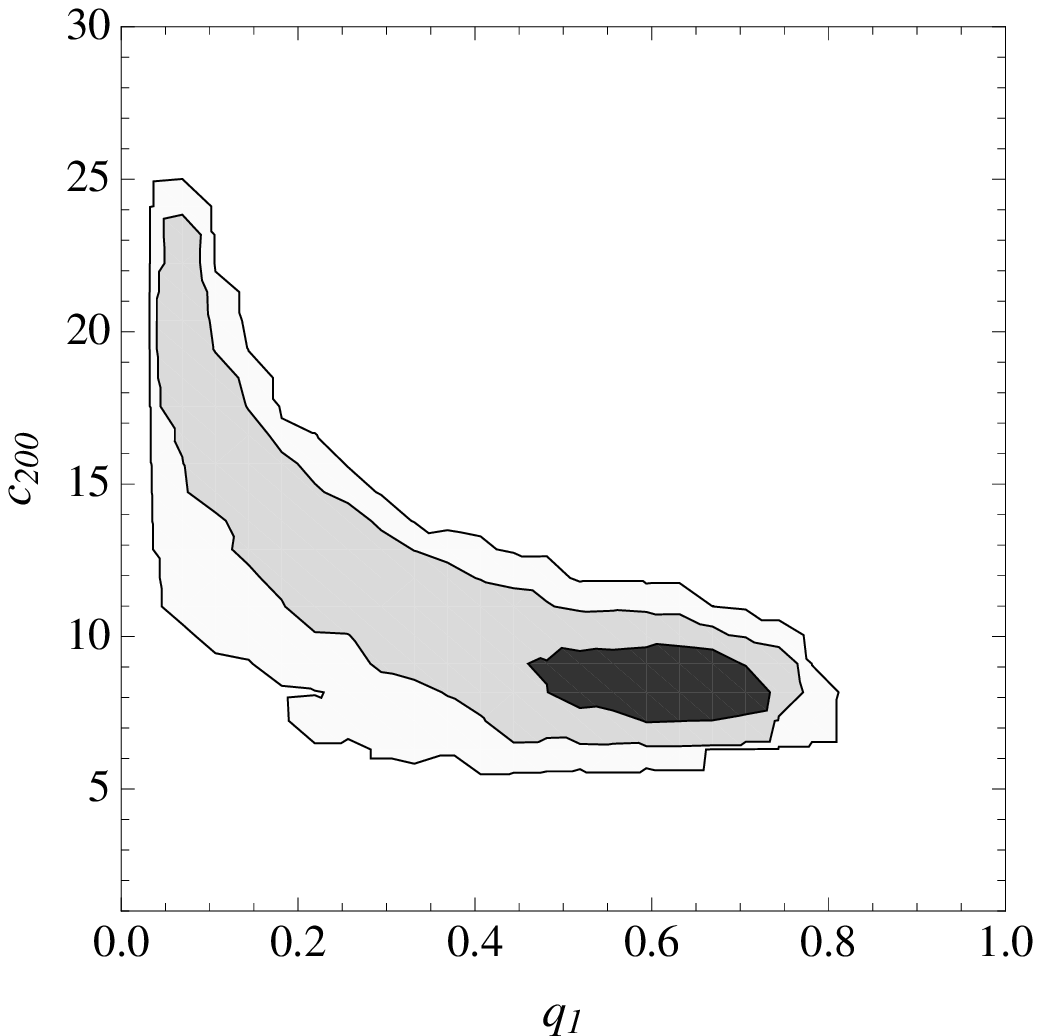}  \\
\end{tabular}
$
\end{center}
  \caption{Contour plot of the marginalised PDF for $q_1$ and $c_{200}$ as derived under the prior assumptions of flat $q$-distribution and random orientation angles. Shadowed contours and panels are as in Fig.~\ref{fig_M200_c200_flat_random_noprior}.}
  \label{fig_q1_c200_nbody_random_noprior}
	\end{figure*}

\begin{figure*}
\begin{center}
$
\begin{tabular}{c@{\hspace{.2cm}}c@{\hspace{.2cm}}c}
   Weak lensing		&Strong lensing	&Weak and Strong lensing \\
   \noalign{\smallskip}
\includegraphics[width=5cm]{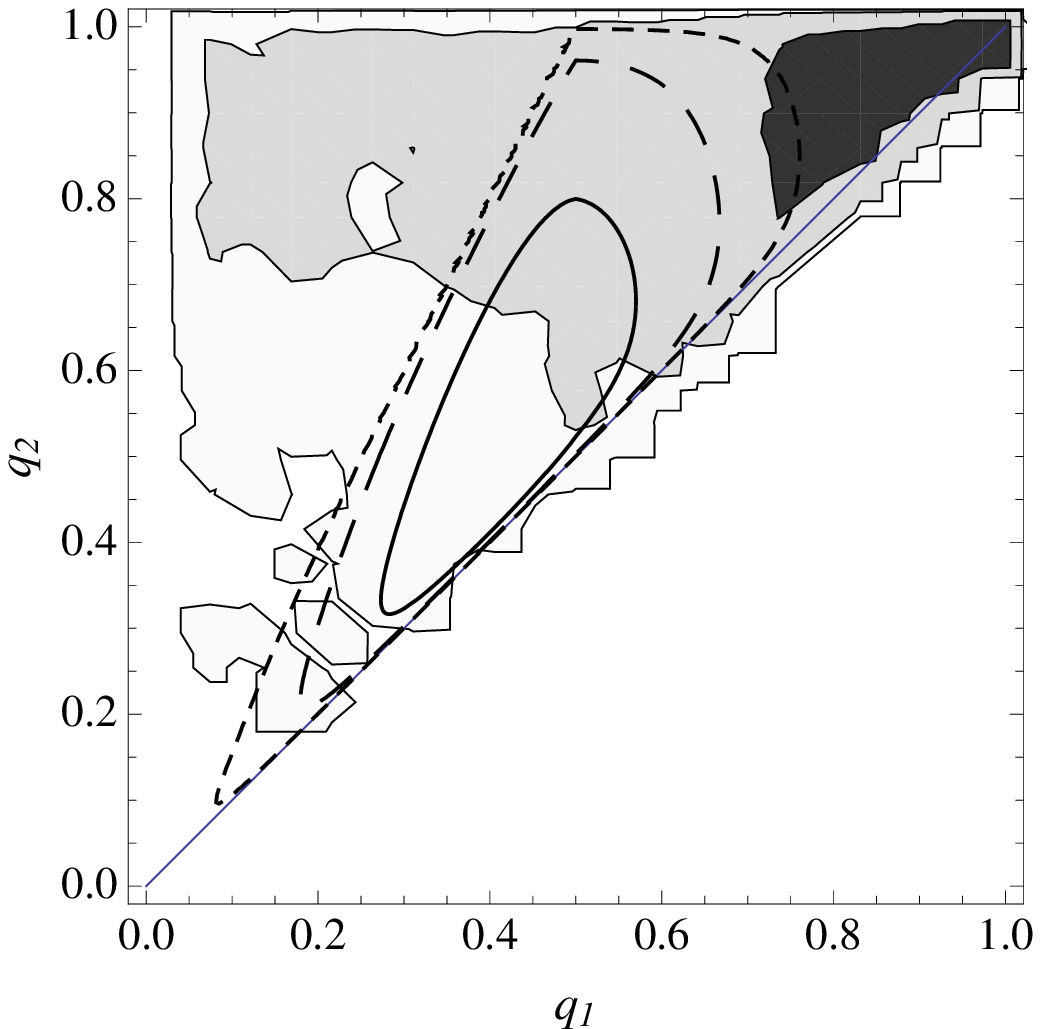} &
\includegraphics[width=5cm]{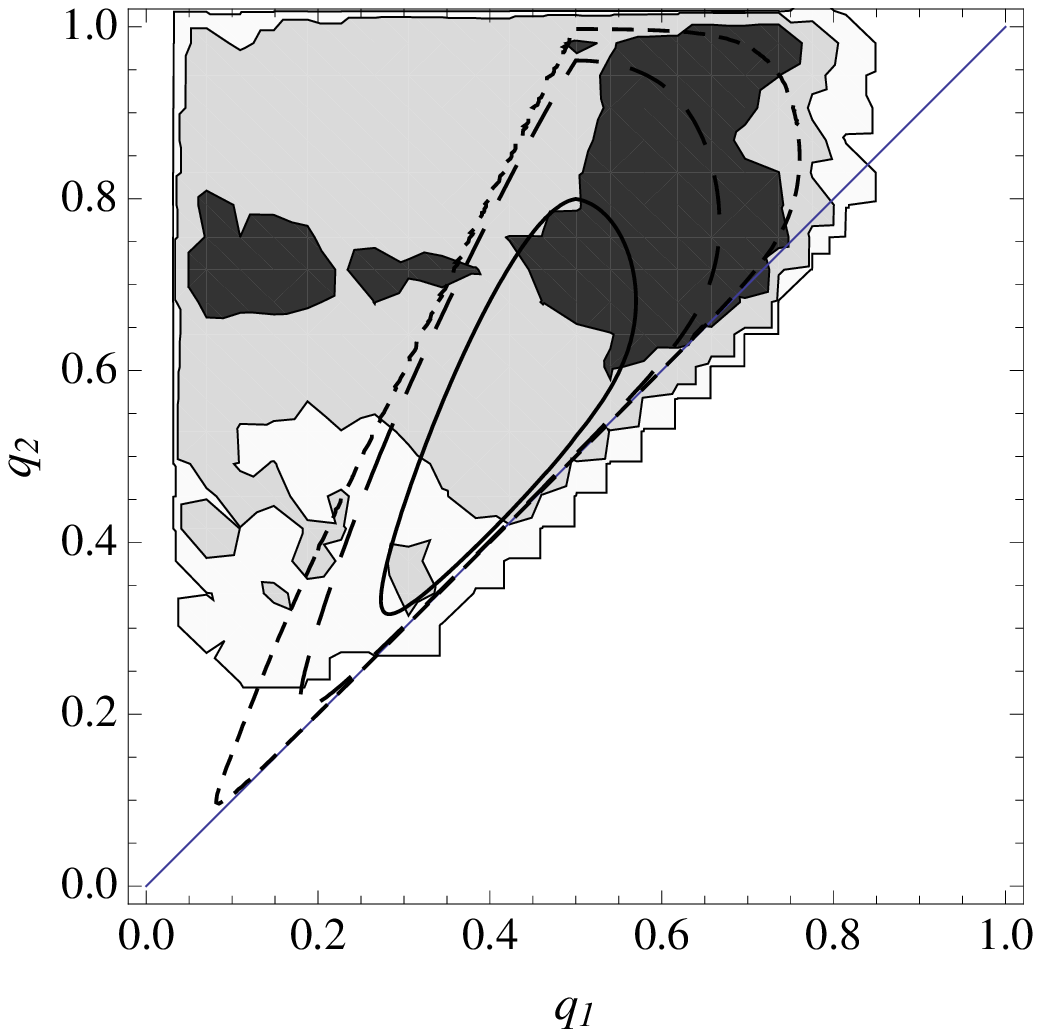} &
\includegraphics[width=5cm]{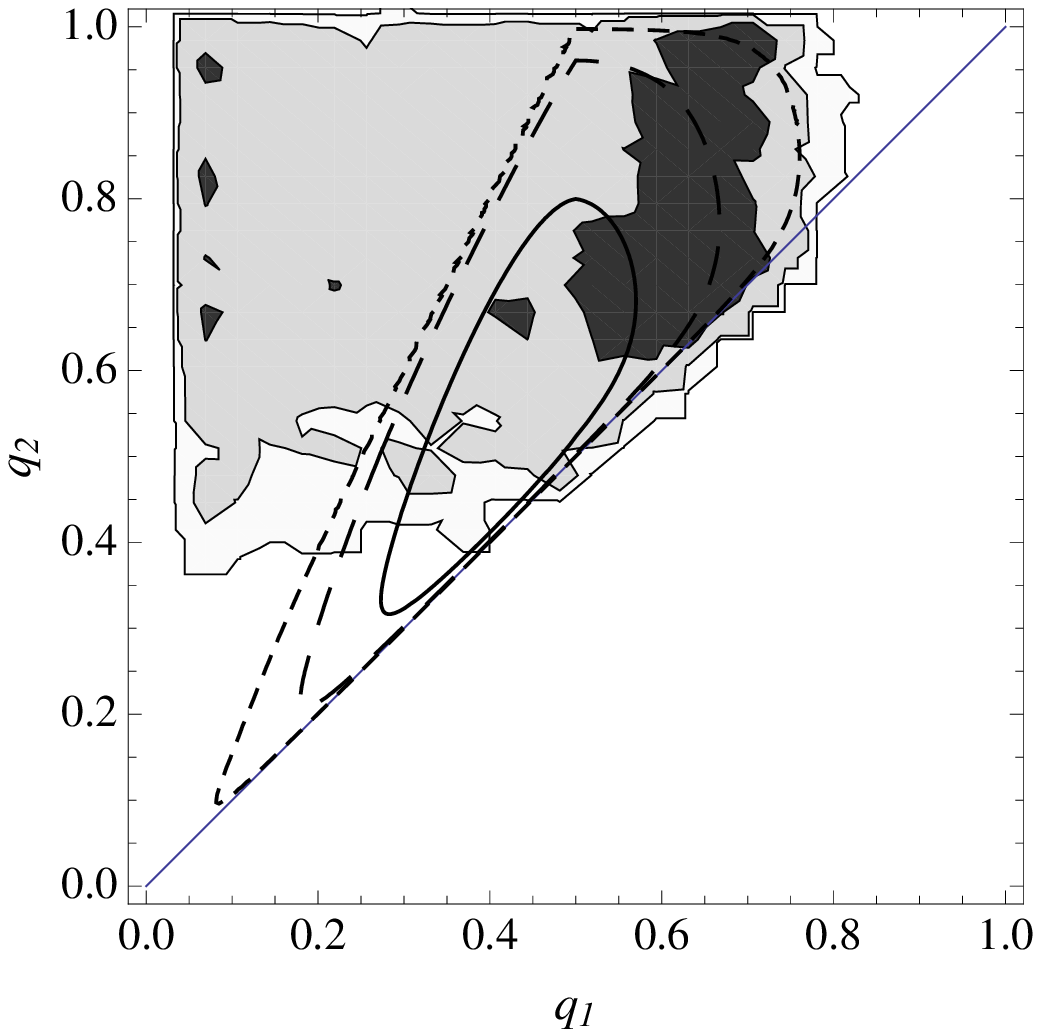}  \\
\end{tabular}
$
\end{center}
  \caption{Contour plot of the marginalised PDF $p(q_1,q_2)$ as derived under the prior assumptions of flat $q$-distribution and random orientation angles. Shadowed contours and panels are as in Fig.~\ref{fig_M200_c200_flat_random_noprior}. The thick full, long dashed and dashed lines limit the 1, 2 and $3\sigma$ confidence regions for the $N$-body like expected distributions.}
  \label{fig_q1_q2_flat_random_noprior}
	\end{figure*}

Even if lensing constraints are not enough to precisely determine the intrinsic axial ratios, mildly triaxial halos with $q_1 \sim 0.5$ make a better job in fitting data than nearly spherical lenses. To see what can be inferred on the shape, let us focus on the results obtained using the less informative priors, i.e. flat distributions for the axial ratios and random orientations. In Fig.~\ref{fig_q1_c200_nbody_random_noprior}, we plotted the marginalised posterior PDF $p (q_1,c_{200})$ under such hypotheses. Values of $0.4 \ls q_1 \ls 0.8$ are more likely than either extremely triaxial or nearly spherical shapes. Results from either weak or strong lensing inversion show correlation between shape and concentration. For intermediate $q_1$ values, the dispersion in $c_{200}$ is quite small.

Triaxial shapes predicted by $N$-body simulations are in good agreement with our results. In Fig.~\ref{fig_q1_q2_flat_random_noprior}, we compared shapes inferred from data with theoretical predictions. We used the flat prior. Even if a priori each configuration $ q_1$ is equally likely, the posterior PDF prefers mildly triaxial axial ratios. Due to the better determination of the projected ellipticity, constraints from strong lensing are more significant. Axial ratios inferred in the weak plus strong lensing case are compatible at $1\sigma$ confidence level with predictions from $N$-body simulations. Furthermore, we can exclude nearly spherical shapes ($q_1 \sim q_2 \sim 1$) at the $3\sigma$ confidence level.

\subsection{Orientation}

\begin{figure*}
\begin{center}
$
\begin{tabular}{c@{\hspace{.2cm}}c@{\hspace{.2cm}}c}
   Weak lensing		&Strong lensing	&Weak and Strong lensing \\
   \noalign{\smallskip}
\includegraphics[width=5cm]{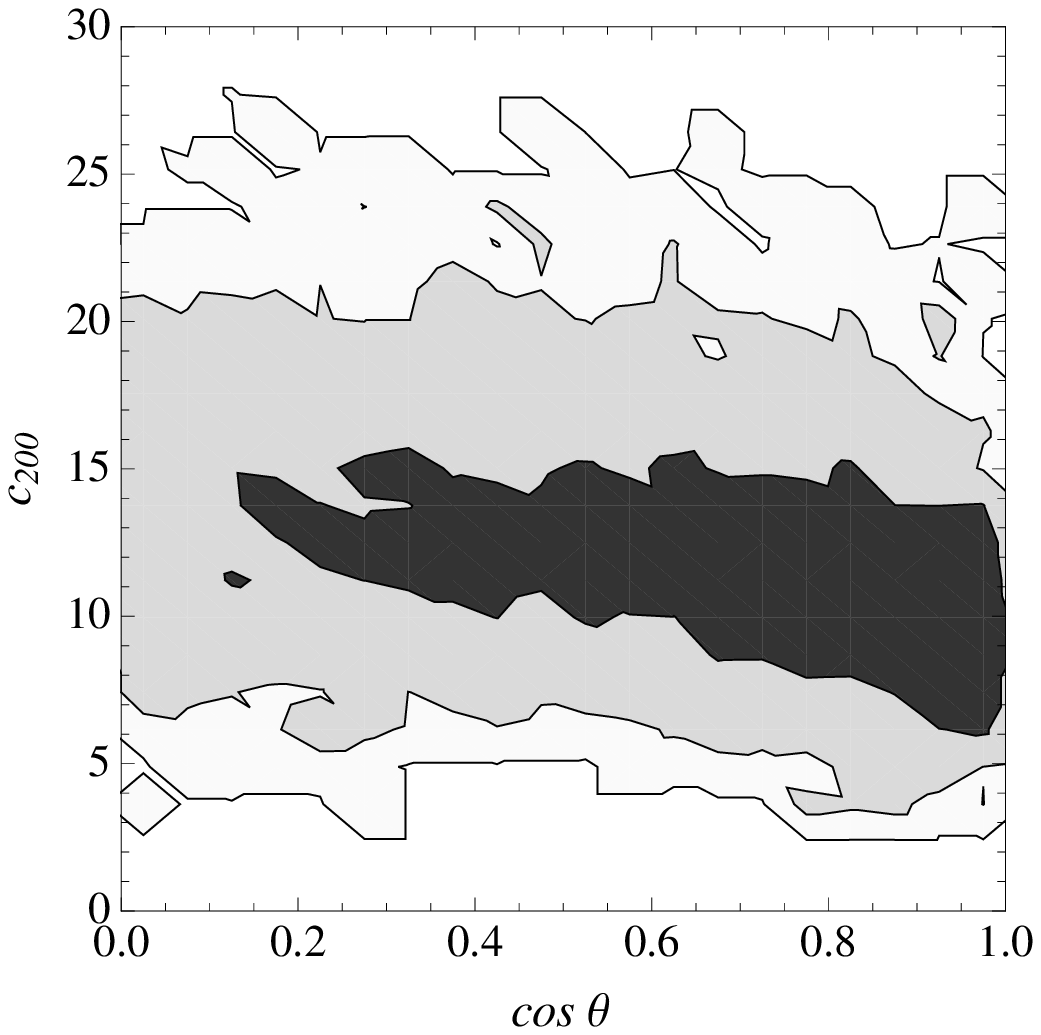} &
\includegraphics[width=5cm]{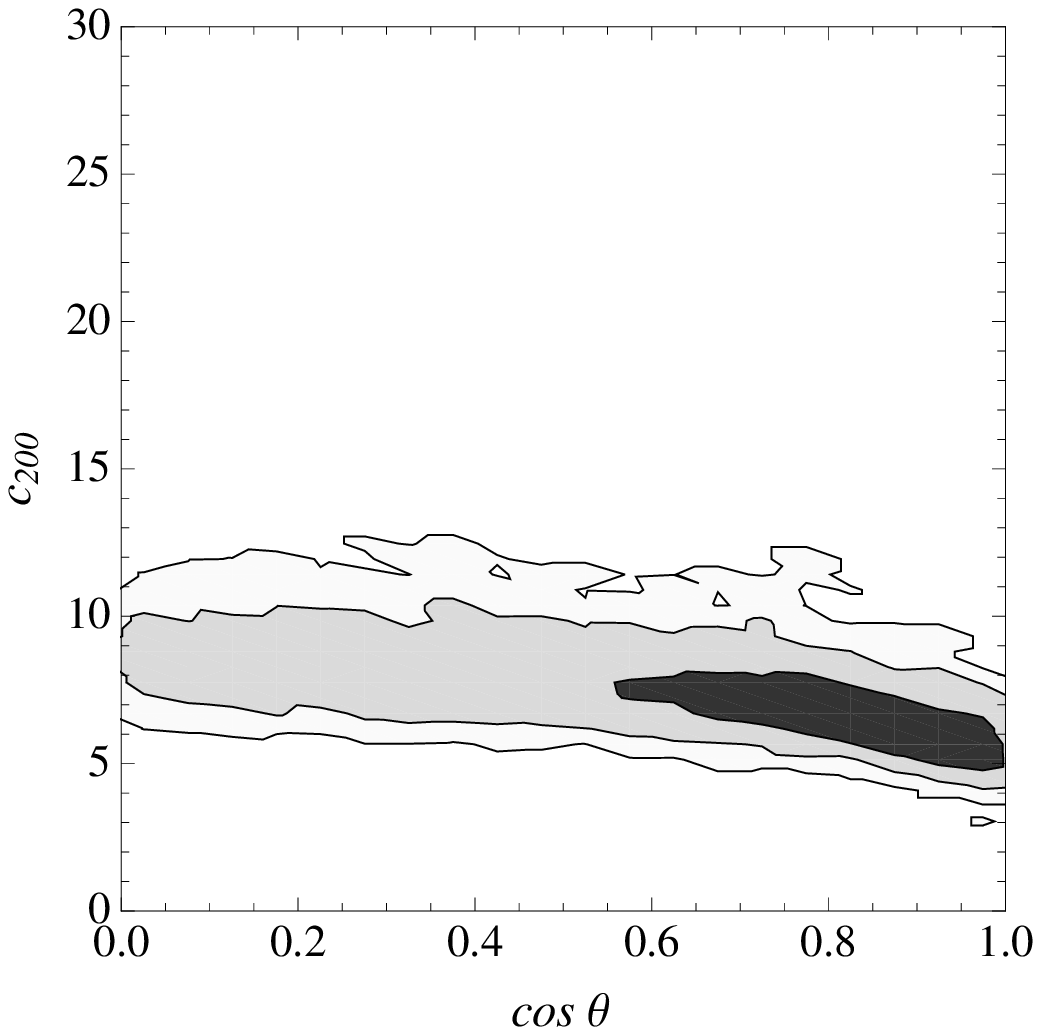} &
\includegraphics[width=5cm]{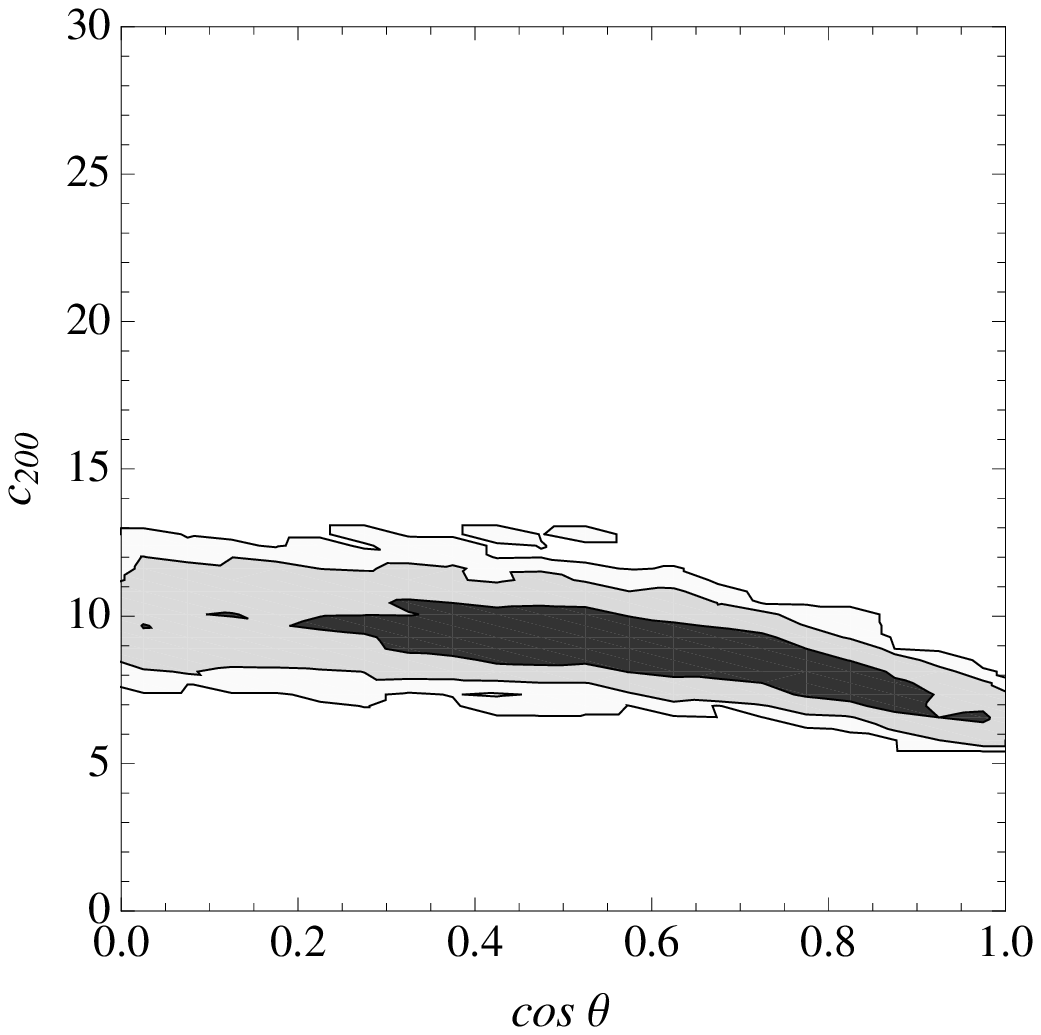}  \\
\end{tabular}
$
\end{center}
  \caption{Contour plot of the marginalised PDF for $\cos \theta$ and $c_{200}$ as derived under the prior assumptions of a $N$-body-like $q$-distribution and random orientation angles. Shadowed contours and panels are as in Fig.~\ref{fig_M200_c200_flat_random_noprior}.}
  \label{fig_costheta_c200_nbody_random_noprior}
	\end{figure*}

There are some indications for an orientations bias. Smaller values of $\theta$ are preferred, pointing out to the cluster being somewhat elongated along the line of sight. This can be seen in Fig.~\ref{fig_costheta_c200_nbody_random_noprior}, where the marginalised $p(\cos \theta, c_{200})$ is plotted under the assumptions of an $N$-body like $q$-distribution but for random orientations. Whereas the tail is extended and small values of $\cos \theta$ are still compatible, the peak of the distribution is at large values of $\cos \theta$. The larger the elongation bias, the smaller the values of the concentration compatible with data. 

We retrieved such trend for either weak (left panel in  in Fig.~\ref{fig_costheta_c200_nbody_random_noprior}) and strong lensing (central panel) inversion. By combining weak to strong lensing data, the preferred values of the concentration shift towards larger values than in the only SL case, but the preference for small inclinations of the major axis, i.e., small values of $\theta$, remains unaffected.

\subsection{Elongation}

\begin{table}
\begin{tabular}[c]{cccr@{$\,\pm\,$}lr@{$\,\pm\,$}lr@{$\,\pm\,$}l}
        \hline
        \noalign{\smallskip}
	\multicolumn{3}{c}{Priors} &\multicolumn{6}{c}{$e_\Delta$} \\ 
        \noalign{\smallskip}
        $q_i$	&angles	&$c(M)$	& \multicolumn{2}{c}{WL} & \multicolumn{2}{c}{SL}	& \multicolumn{2}{c}{WL+SL}	 \\
        \hline
         \noalign{\smallskip}
	flat		&random	&$\times$			&$1.1$	&$0.3$	&$1.5$	&$0.7$	&$1.5$	&$0.7$ \\
	flat		&bias	&$\times$			&$0.9$	&$0.1$	&$0.9$	&$0.2$	&$1.0$	&$0.2$ \\
	flat		&random	&$\checkmark$		&$1.1$	&$0.2$	&$1.1$	&$0.2$	&$1.0$	&$0.3$ \\
	\hline
	$N$-body	&random	&$\times$			&$1.3$	&$0.5$	&$1.2$	&$0.4$	&$1.5$	&$0.4$ \\
	$N$-body	&bias	&$\times$			&$0.8$	&$0.2$	&$0.8$	&$0.2$	&$0.9$	&$0.2$ \\
	$N$-body	&bias	&$\checkmark$		&$0.7$	&$0.2$	&$0.7$	&$0.2$	&$0.8$	&$0.2$ \\
    	\hline
\end{tabular}
\caption{Elongation parameter inferred by assuming different priors. Biweight estimators are reported.}
\label{tab_elongation}
\end{table}


The cluster length along the line of sight can be parametrized by the elongation $e_\Delta$ \citep{ser07}. With no regard to either the prior assumptions or the data-set employed (either strong or weak lensing), results are quite consistent, see Table~\ref{tab_elongation}. Very elongated lenses ($e_\Delta \ll 1$) may be quite strong lenses. Then, when we enforce a biased orientation, clusters turn out to be less massive and concentrated since lensing strength is provided by the elongation. 

For halos with an orientation bias and intrinsic ratios as from $N$-body simulations, the line of sight elongation turns out to be $\sim 20-30$ per cent larger than the projected dimension in the plane of the sky. However, due to the large statistical uncertainty, more spherical configurations are compatible too. When using triaxial axis ratios as inferred from $N$-body simulations, an unrealistic tail at very large values for the PDF of the elongation is cut off. In the case of a flat prior for the $q_i$, this tail pushes the central estimate for the elongation to higher values.

These new results from lensing for the matter halo agree with a previous direct measurement for the gas distribution ($e_\Delta = 1.1 \pm 0.2$) obtained by combining X-ray and SZ data \citep{def+al05,ser+al06}. The matter distribution is expected to be more triaxial than the intra-cluster medium, but the large errors do not allow us to confirm this prediction.

Recently, \citet{mor+al11} combined lensing and X-ray data to infer a quite elongated halo. Assuming the cluster to be aligned with the line of sight and a peculiar modelling for the deviation from hydrostatic equilibrium, \citet{mor+al11} calculated $e_\Delta \sim 0.6$. This estimate strongly relies on the hypothesis on the hydrostatic status and on results from numerical simulations relating the shapes of gravitational potential and the matter distribution, which are affected by quite large errors for very elongated clusters \citep{lee+sut03}.

\section{Conclusions}
\label{sec_conc}

Most of the tension on A1689 as an extremely over-concentrated lensing cluster can be eliminated. We exploited two main ingredients: $i)$ a full triaxial investigation of the lensing halo; $ii)$ a Bayesian statistical analysis. Simulations and theoretical analyses show that massive halos are quite triaxial. A rich variety of shapes and orientations has to be considered, some of which allow moderately concentrated halos to be lenses as effective as much more intrinsically concentrated clusters. In particular, lensing cross-section for ellipsoidal halos elongated along the line of sight are enhanced.

Bayesian methods allow us to explore the full parameter space, avoiding to restrict the analysis in the vicinity of the configuration with the maximum likelihood. A proper investigation of the parameter space provides also a more reliable estimate for the statistical uncertainty.

Our results are in good agreement with predictions based on numerical simulations. The triaxial lensing analysis of A1689 suggests a mildly triaxial cluster. Computed concentrations ($c_{200}=10\pm3 $ or $7\pm1$ for the only WL or the combined WL+SL case) are only slightly larger ($\ls 2\sigma$) than the median concentration ($c_{200}\simeq 3.2 \pm 1.6$) expected for the cluster mass $M_{200}=13\times10^{14}M_\odot$. Inversions of either strong or weak lensing data are reassuringly compatible, and the combined analysis enhanced the picture of A1689 as a typical massive lensing cluster. There is some evidence for an alignment of the major axis with the line of sight. The size of the matter distribution along the line of sight is larger by $\sim 20$ per cent than the width in the plane of the sky, but more spherical configurations are still compatible with data.

The theoretical form of the $c(M)$ relation is still debated. Recently, \citet{pra+al11} showed that the dependence of concentration on halo mass and its evolution can be obtained from the root-mean-square (rms) fluctuation amplitude of the linear density field $\sigma(M,z)$. They noticed a flattening and upturn of the relation with increasing mass. Estimated concentrations for galaxy clusters might be substantially larger than previous results. Agreement with our result for A1689 would consequently be even better, since they predict a median $c_{200}=5.1$ for a cluster of $M_{200}=13\times10^{14}M_\odot$ at the redshift of A1689. Accurate determinations of $n_s$, the spectral index of the primordial power spectrum, and $\sigma_8$, the rms amplitude of linear mass fluctuations, are also needed for reliable theoretical prediction on $c(M)$.

Due to the different statistical approaches, it is difficult to compare our estimates with previous works. Our analysis is in broad agreement with previous results which found a concentration of $c_{200}\sim 6$ for strong lensing analyses or $\sim 8$--$12$ for weak lensing analyses. Previous results from literature have already been collected in \citet[ table~4]{cor+al09} and  \citet[ table~2]{coe+al10}.  In general, triaxiality allows for smaller concentrations and larger dispersions. Assuming spherical symmetry, \citet{ume+al09} found $c_{200}= 12.5^{+3.9}_{-2.7}$ from their one-dimensional WL analysis of Subaru distortion data. Due to the strict a priori hypothesis on the intrinsic shape, this estimate is higher than what obtained in the present paper allowing for triaxiality. The dispersion for the spherical symmetric case is smaller than our WL result assuming a flat prior on axial ratios and random orientations. Forcing the shape to follow results from $N$-body simulations and the orientation to be biased, dispersion is comparable with the spherical symmetric case, but the estimated concentration is significantly lower ($c_{200} \sim 10$ vs. $c_{200} \sim 13$). 

The most meaningful comparison can be made with \citet{cor+al09}, who first employed a Bayesian analysis to deproject weak lensing data of A1689\footnote{The ``flat'' model in \citet{cor+al09} used priors similar to our flat $q$-distribution with random orientations; their ``Shaw'' model used priors for the axial ratios from numerical simulations.}. Even if they used a different data set obtained with the ESO/MPG Wide Field Imager, the obtained posterior distributions for $c_{200}$ are in agreement with ours for both location and dispersion.

Orientation and shape biases can be very sizable when estimating halo concentrations from lensing but are less effective in X-ray analyses \citep{og+bl09,men+al10}, which usually provide lesser values for $c_{200}$ in A1689 \citep{coe+al10}. A proper statistical modelling of triaxiality is then mandatory for reliable estimates. This deserves attention since an unbiased lensing estimate can provide concentration values without assuming hydrostatic equilibrium, which can plague X-ray methods \citep{mol+al10}. Combined X-ray and SZ methods are interesting too, since they can directly infer the elongation of the intra-cluster medium distribution without the hydrostatic equilibrium hypothesis \citep{def+al05,ser+al06,ser07}. However, the gas density is expected to differ from the dark matter profile. Some further hypotheses have to be used to link the gas to the dark matter distribution.

\section*{Acknowledgements}
MS thanks M. Limousin for making available some tabulated results of the strong lensing analysis in \citet{lim+al07}. KU acknowledges support from the Academia Sinica Career Development Award and the  National Science Council of Taiwan under the grant NSC97- 2112-M-001-020-MY3.


\begin{thebibliography}{}

\setlength{\itemindent}{-2.5em}


\bibitem[\protect\citeauthoryear{{Beers}, {Flynn} \& {Gebhardt}}{{Beers}
  et~al.}{1990}]{bee+al90}
{Beers} T.~C.,  {Flynn} K.,    {Gebhardt} K.,  1990, \aj, 100, 32

\bibitem[\protect\citeauthoryear{{Binggeli}}{{Binggeli}}{1980}]{bin80}
{Binggeli} B.,  1980, A\&A, 82, 289

\bibitem[\protect\citeauthoryear{{Broadhurst}, {Ben{\'{\i}}tez}, {Coe},
  {Sharon}, {Zekser}, {White}, {Ford}, {Bouwens}, {Blakeslee}, {Clampin},
  {Cross} \& {et al.}}{{Broadhurst} et~al.}{2005}]{bro+al05}
{Broadhurst} T.,  {Ben{\'{\i}}tez} N.,  {Coe} D.,  {Sharon} K.,  {Zekser} K.,
  {White} R.,  {Ford} H.,  {Bouwens} R.,  {Blakeslee} J.,  {Clampin} M.,
  {Cross} N.,    {et al.} 2005, \apj, 621, 53

\bibitem[\protect\citeauthoryear{{Broadhurst}, {Takada}, {Umetsu}, {Kong},
  {Arimoto}, {Chiba} \& {Futamase}}{{Broadhurst} et~al.}{2005}]{bro+al05b}
{Broadhurst} T.,  {Takada} M.,  {Umetsu} K.,  {Kong} X.,  {Arimoto} N.,
  {Chiba} M.,    {Futamase} T.,  2005, \apjl, 619, L143

\bibitem[\protect\citeauthoryear{{Broadhurst}, {Umetsu}, {Medezinski}, {Oguri}
  \& {Rephaeli}}{{Broadhurst} et~al.}{2008}]{bro+al08}
{Broadhurst} T.,  {Umetsu} K.,  {Medezinski} E.,  {Oguri} M.,    {Rephaeli} Y.,
   2008, \apjl, 685, L9

\bibitem[\protect\citeauthoryear{{Broadhurst} \& {Barkana}}{{Broadhurst} \&
  {Barkana}}{2008}]{br+ba08}
{Broadhurst} T.~J.,  {Barkana} R.,  2008, \mnras, 390, 1647

\bibitem[\protect\citeauthoryear{{Broadhurst}, {Taylor} \&
  {Peacock}}{{Broadhurst} et~al.}{1995}]{bro+al95}
{Broadhurst} T.~J.,  {Taylor} A.~N.,    {Peacock} J.~A.,  1995, \apj, 438, 49

\bibitem[\protect\citeauthoryear{{Bullock}, {Kolatt}, {Sigad}, {Somerville},
  {Kravtsov}, {Klypin}, {Primack} \& {Dekel}}{{Bullock}
  et~al.}{2001}]{bul+al01}
{Bullock} J.~S.,  {Kolatt} T.~S.,  {Sigad} Y.,  {Somerville} R.~S.,  {Kravtsov}
  A.~V.,  {Klypin} A.~A.,  {Primack} J.~R.,    {Dekel} A.,  2001, \mnras, 321,
  559

\bibitem[\protect\citeauthoryear{{Coe}, {Benitez}, {Broadhurst}, {Moustakas} \&
  {Ford}}{{Coe} et~al.}{2010}]{coe+al10}
{Coe} D.,  {Benitez} N.,  {Broadhurst} T.,  {Moustakas} L.,    {Ford} H.,
  2010, ArXiv: 1005.0398

\bibitem[\protect\citeauthoryear{{Comerford} \& {Natarajan}}{{Comerford} \&
  {Natarajan}}{2007}]{co+na07}
{Comerford} J.~M.,  {Natarajan} P.,  2007, \mnras, 379, 190

\bibitem[\protect\citeauthoryear{{Corless} \& {King}}{{Corless} \&
  {King}}{2007}]{co+ki07}
{Corless} V.~L.,  {King} L.~J.,  2007, \mnras, 380, 149

\bibitem[\protect\citeauthoryear{{Corless}, {King} \& {Clowe}}{{Corless}
  et~al.}{2009}]{cor+al09}
{Corless} V.~L.,  {King} L.~J.,    {Clowe} D.,  2009, \mnras, 393, 1235

\bibitem[\protect\citeauthoryear{{De Filippis}, {Sereno}, {Bautz} \&
  {Longo}}{{De Filippis} et~al.}{2005}]{def+al05}
{De Filippis} E.,  {Sereno} M.,  {Bautz} M.~W.,    {Longo} G.,  2005, \apj,
  625, 108

\bibitem[\protect\citeauthoryear{{Diemand}, {Moore} \& {Stadel}}{{Diemand}
  et~al.}{2004}]{die+al04}
{Diemand} J.,  {Moore} B.,    {Stadel} J.,  2004, \mnras, 353, 624

\bibitem[\protect\citeauthoryear{{Duffy}, {Schaye}, {Kay} \& {Dalla
  Vecchia}}{{Duffy} et~al.}{2008}]{duf+al08}
{Duffy} A.~R.,  {Schaye} J.,  {Kay} S.~T.,    {Dalla Vecchia} C.,  2008,
  \mnras, 390, L64

\bibitem[\protect\citeauthoryear{{Gao}, {Navarro}, {Cole}, {Frenk}, {White},
  {Springel}, {Jenkins} \& {Neto}}{{Gao} et~al.}{2008}]{gao+al08}
{Gao} L.,  {Navarro} J.~F.,  {Cole} S.,  {Frenk} C.~S.,  {White} S.~D.~M.,
  {Springel} V.,  {Jenkins} A.,    {Neto} A.~F.,  2008, \mnras, 387, 536

\bibitem[\protect\citeauthoryear{{Halkola}, {Seitz} \& {Pannella}}{{Halkola}
  et~al.}{2006}]{hal+al06}
{Halkola} A.,  {Seitz} S.,    {Pannella} M.,  2006, \mnras, 372, 1425

\bibitem[\protect\citeauthoryear{{Hamana}, {Miyazaki}, {Kashikawa}, {Ellis},
  {Massey}, {Refregier} \& {Taylor}}{{Hamana} et~al.}{2009}]{ham+al09}
{Hamana} T.,  {Miyazaki} S.,  {Kashikawa} N.,  {Ellis} R.~S.,  {Massey} R.~J.,
  {Refregier} A.,    {Taylor} J.~E.,  2009, \pasj, 61, 833

\bibitem[\protect\citeauthoryear{{Hennawi}, {Dalal}, {Bode} \&
  {Ostriker}}{{Hennawi} et~al.}{2007}]{hen+al07}
{Hennawi} J.~F.,  {Dalal} N.,  {Bode} P.,    {Ostriker} J.~P.,  2007, \apj,
  654, 714

\bibitem[\protect\citeauthoryear{{Jing} \& {Suto}}{{Jing} \&
  {Suto}}{2002}]{ji+su02}
{Jing} Y.~P.,  {Suto} Y.,  2002, \apj, 574, 538

\bibitem[\protect\citeauthoryear{{Johnston}, {Sheldon}, {Wechsler}, {Rozo},
  {Koester}, {Frieman}, {McKay}, {Evrard}, {Becker} \& {Annis}}{{Johnston}
  et~al.}{2007}]{joh+al07}
{Johnston} D.~E.,  {Sheldon} E.~S.,  {Wechsler} R.~H.,  {Rozo} E.,  {Koester}
  B.~P.,  {Frieman} J.~A.,  {McKay} T.~A.,  {Evrard} A.~E.,  {Becker} M.~R.,
  {Annis} J.,  2007, ArXiv: 0709.1159

\bibitem[\protect\citeauthoryear{{Kawahara}, {Suto}, {Kitayama}, {Sasaki},
  {Shimizu}, {Rasia} \& {Dolag}}{{Kawahara} et~al.}{2007}]{kaw+al07}
{Kawahara} H.,  {Suto} Y.,  {Kitayama} T.,  {Sasaki} S.,  {Shimizu} M.,
  {Rasia} E.,    {Dolag} K.,  2007, \apj, 659, 257

\bibitem[\protect\citeauthoryear{{Kawaharada}, {Okabe}, {Umetsu}, {Takizawa},
  {Matsushita}, {Fukazawa}, {Hamana}, {Miyazaki}, {Nakazawa} \&
  {Ohashi}}{{Kawaharada} et~al.}{2010}]{kaw+al10}
{Kawaharada} M.,  {Okabe} N.,  {Umetsu} K.,  {Takizawa} M.,  {Matsushita} K.,
  {Fukazawa} Y.,  {Hamana} T.,  {Miyazaki} S.,  {Nakazawa} K.,    {Ohashi} T.,
  2010, \apj, 714, 423

\bibitem[\protect\citeauthoryear{{Keeton}}{{Keeton}}{2001a}]{kee01b}
{Keeton} C.~R.,  2001a, astro-ph/0102341

\bibitem[\protect\citeauthoryear{{Keeton}}{{Keeton}}{2001b}]{kee01a}
{Keeton} C.~R.,  2001b, astro-ph/0102340

\bibitem[\protect\citeauthoryear{{Lau}, {Kravtsov} \& {Nagai}}{{Lau}
  et~al.}{2009}]{lau+al09}
{Lau} E.~T.,  {Kravtsov} A.~V.,    {Nagai} D.,  2009, \apj, 705, 1129

\bibitem[\protect\citeauthoryear{{Lee} \& {Suto}}{{Lee} \&
  {Suto}}{2003}]{lee+sut03}
{Lee} J.,  {Suto} Y.,  2003, \apj, 585, 151

\bibitem[\protect\citeauthoryear{{Lemze}, {Broadhurst}, {Rephaeli}, {Barkana}
  \& {Umetsu}}{{Lemze} et~al.}{2009}]{lem+al08}
{Lemze} D.,  {Broadhurst} T.,  {Rephaeli} Y.,  {Barkana} R.,    {Umetsu} K.,
  2009, \apj, 701, 1336

\bibitem[\protect\citeauthoryear{{Limousin}, {Richard}, {Jullo}, {Kneib},
  {Fort}, {Soucail}, {El{\'{\i}}asd{\'o}ttir}, {Natarajan}, {Ellis}, {Smail},
  {Czoske}, {Smith}, {Hudelot}, {Bardeau}, {Ebeling}, {Egami} \&
  {Knudsen}}{{Limousin} et~al.}{2007}]{lim+al07}
{Limousin} M.,  {Richard} J.,  {Jullo} E.,  {Kneib} J.-P.,  {Fort} B.,
  {Soucail} G.,  {El{\'{\i}}asd{\'o}ttir} {\'A}.,  {Natarajan} P.,  {Ellis}
  R.~S.,  {Smail} I.,  {Czoske} O.,  {Smith} G.~P.,  {Hudelot} P.,  {Bardeau}
  S.,  {Ebeling} H.,  {Egami} E.,    {Knudsen} K.~K.,  2007, \apj, 668, 643

\bibitem[\protect\citeauthoryear{{Macci{\`o}}, {Dutton} \& {van den
  Bosch}}{{Macci{\`o}} et~al.}{2008}]{mac+al08}
{Macci{\`o}} A.~V.,  {Dutton} A.~A.,    {van den Bosch} F.~C.,  2008, \mnras,
  391, 1940

\bibitem[\protect\citeauthoryear{{Mandelbaum}, {Seljak} \&
  {Hirata}}{{Mandelbaum} et~al.}{2008}]{man+al08}
{Mandelbaum} R.,  {Seljak} U.,    {Hirata} C.~M.,  2008, JCAP, 8, 6

\bibitem[\protect\citeauthoryear{{Meneghetti}, {Fedeli}, {Pace},
  {Gottl{\"o}ber} \& {Yepes}}{{Meneghetti} et~al.}{2010}]{men+al10}
{Meneghetti} M.,  {Fedeli} C.,  {Pace} F.,  {Gottl{\"o}ber} S.,    {Yepes} G.,
  2010, \aap, 519, A90

\bibitem[\protect\citeauthoryear{{Meneghetti}, {Fedeli}, {Zitrin},
  {Bartelmann}, {Broadhurst}, {Gottl{\"o}ber}, {Moscardini} \&
  {Yepes}}{{Meneghetti} et~al.}{2011}]{men+al11}
{Meneghetti} M.,  {Fedeli} C.,  {Zitrin} A.,  {Bartelmann} M.,  {Broadhurst}
  T.,  {Gottl{\"o}ber} S.,  {Moscardini} L.,    {Yepes} G.,  2011, \aap, 530,
  A17

\bibitem[\protect\citeauthoryear{{Miyazaki}, {Hamana}, {Shimasaku}, {Furusawa},
  {Doi}, {Hamabe}, {Imi}, {Kimura}, {Komiyama}, {Nakata}, {Okada}, {Okamura},
  {Ouchi}, {Sekiguchi}, {Yagi} \& {Yasuda}}{{Miyazaki} et~al.}{2002}]{miy+al02}
{Miyazaki} S.,  {Hamana} T.,  {Shimasaku} K.,  {Furusawa} H.,  {Doi} M.,
  {Hamabe} M.,  {Imi} K.,  {Kimura} M.,  {Komiyama} Y.,  {Nakata} F.,  {Okada}
  N.,  {Okamura} S.,  {Ouchi} M.,  {Sekiguchi} M.,  {Yagi} M.,    {Yasuda} N.,
  2002, \apjl, 580, L97

\bibitem[\protect\citeauthoryear{{Molnar}, {Chiu}, {Umetsu}, {Chen}, {Hearn},
  {Broadhurst}, {Bryan} \& {Shang}}{{Molnar} et~al.}{2010}]{mol+al10}
{Molnar} S.~M.,  {Chiu} I.-N.,  {Umetsu} K.,  {Chen} P.,  {Hearn} N.,
  {Broadhurst} T.,  {Bryan} G.,    {Shang} C.,  2010, \apjl, 724, L1

\bibitem[\protect\citeauthoryear{{Morandi}, {Limousin}, {Rephaeli}, {Umetsu},
  {Barkana}, {Broadhurst} \& {Dahle}}{{Morandi} et~al.}{2011}]{mor+al11}
{Morandi} A.,  {Limousin} M.,  {Rephaeli} Y.,  {Umetsu} K.,  {Barkana} R.,
  {Broadhurst} T.,    {Dahle} H.,  2011, ArXiv: 1103.0202

\bibitem[\protect\citeauthoryear{{Navarro}, {Frenk} \& {White}}{{Navarro}
  et~al.}{1996}]{nfw96}
{Navarro} J.~F.,  {Frenk} C.~S.,    {White} S.~D.~M.,  1996, \apj, 462, 563

\bibitem[\protect\citeauthoryear{{Navarro}, {Frenk} \& {White}}{{Navarro}
  et~al.}{1997}]{nav+al97}
{Navarro} J.~F.,  {Frenk} C.~S.,    {White} S.~D.~M.,  1997, \apj, 490, 493

\bibitem[\protect\citeauthoryear{{Neto}, {Gao}, {Bett}, {Cole}, {Navarro},
  {Frenk}, {White}, {Springel} \& {Jenkins}}{{Neto} et~al.}{2007}]{net+al07}
{Neto} A.~F.,  {Gao} L.,  {Bett} P.,  {Cole} S.,  {Navarro} J.~F.,  {Frenk}
  C.~S.,  {White} S.~D.~M.,  {Springel} V.,    {Jenkins} A.,  2007, \mnras,
  381, 1450

\bibitem[\protect\citeauthoryear{{Oguri} \& {Blandford}}{{Oguri} \&
  {Blandford}}{2009}]{og+bl09}
{Oguri} M.,  {Blandford} R.~D.,  2009, \mnras, 392, 930

\bibitem[\protect\citeauthoryear{{Oguri}, {Hennawi}, {Gladders}, {Dahle},
  {Natarajan}, {Dalal}, {Koester}, {Sharon} \& {Bayliss}}{{Oguri}
  et~al.}{2009}]{ogu+al09}
{Oguri} M.,  {Hennawi} J.~F.,  {Gladders} M.~D.,  {Dahle} H.,  {Natarajan} P.,
  {Dalal} N.,  {Koester} B.~P.,  {Sharon} K.,    {Bayliss} M.,  2009, \apj,
  699, 1038

\bibitem[\protect\citeauthoryear{{Oguri}, {Takada}, {Okabe} \& {Smith}}{{Oguri}
  et~al.}{2010}]{ogu+al10}
{Oguri} M.,  {Takada} M.,  {Okabe} N.,    {Smith} G.~P.,  2010, \mnras, 405,
  2215

\bibitem[\protect\citeauthoryear{{Oguri}, {Takada}, {Umetsu} \&
  {Broadhurst}}{{Oguri} et~al.}{2005}]{ogu+al05}
{Oguri} M.,  {Takada} M.,  {Umetsu} K.,    {Broadhurst} T.,  2005, \apj, 632,
  841

\bibitem[\protect\citeauthoryear{{Okabe}, {Takada}, {Umetsu}, {Futamase} \&
  {Smith}}{{Okabe} et~al.}{2010}]{oka+al10}
{Okabe} N.,  {Takada} M.,  {Umetsu} K.,  {Futamase} T.,    {Smith} G.~P.,
  2010, \pasj, 62, 811

\bibitem[\protect\citeauthoryear{{Peng}, {Andersson}, {Bautz} \&
  {Garmire}}{{Peng} et~al.}{2009}]{pen+al09}
{Peng} E.-H.,  {Andersson} K.,  {Bautz} M.~W.,    {Garmire} G.~P.,  2009, \apj,
  701, 1283

\bibitem[\protect\citeauthoryear{{Prada}, {Klypin}, {Cuesta}, {Betancort-Rijo}
  \& {Primack}}{{Prada} et~al.}{2011}]{pra+al11}
{Prada} F.,  {Klypin} A.~A.,  {Cuesta} A.~J.,  {Betancort-Rijo} J.~E.,
  {Primack} J.,  2011, ArXiv: 1104.5130

\bibitem[\protect\citeauthoryear{{Sadeh} \& {Rephaeli}}{{Sadeh} \&
  {Rephaeli}}{2008}]{sa+re08}
{Sadeh} S.,  {Rephaeli} Y.,  2008, \mnras, 388, 1759

\bibitem[\protect\citeauthoryear{{Sand}, {Treu}, {Ellis}, {Smith} \&
  {Kneib}}{{Sand} et~al.}{2008}]{san+al08}
{Sand} D.~J.,  {Treu} T.,  {Ellis} R.~S.,  {Smith} G.~P.,    {Kneib} J.-P.,
  2008, \apj, 674, 711

\bibitem[\protect\citeauthoryear{{Sereno}}{{Sereno}}{2007}]{ser07}
{Sereno} M.,  2007, \mnras, 380, 1207

\bibitem[\protect\citeauthoryear{{Sereno}, {De Filippis}, {Longo} \&
  {Bautz}}{{Sereno} et~al.}{2006}]{ser+al06}
{Sereno} M.,  {De Filippis} E.,  {Longo} G.,    {Bautz} M.~W.,  2006, \apj,
  645, 170

\bibitem[\protect\citeauthoryear{{Sereno}, {Jetzer} \& {Lubini}}{{Sereno}
  et~al.}{2010}]{ser+al10}
{Sereno} M.,  {Jetzer} P.,    {Lubini} M.,  2010, \mnras, 403, 2077

\bibitem[\protect\citeauthoryear{{Sereno}, {Lubini} \& {Jetzer}}{{Sereno}
  et~al.}{2010}]{ser+al09}
{Sereno} M.,  {Lubini} M.,    {Jetzer} P.,  2010, \aap, 518, A55

\bibitem[\protect\citeauthoryear{{Stark}}{{Stark}}{1977}]{sta77}
{Stark} A.~A.,  1977, \apj, 213, 368

\bibitem[\protect\citeauthoryear{{Umetsu}, {Birkinshaw}, {Liu}, {Wu},
  {Medezinski}, {Broadhurst}, {Lemze}, {Zitrin}, {Ho}, {Huang}, {Koch}, {Liao},
  {Lin}, {Molnar}, {Nishioka}, {Wang} \& {et al.}}{{Umetsu}
  et~al.}{2009}]{ume+al09}
{Umetsu} K.,  {Birkinshaw} M.,  {Liu} G.,  {Wu} J.,  {Medezinski} E.,
  {Broadhurst} T.,  {Lemze} D.,  {Zitrin} A.,  {Ho} P.~T.~P.,  {Huang} C.,
  {Koch} P.~M.,  {Liao} Y.,  {Lin} K.,  {Molnar} S.~M.,  {Nishioka} H.,  {Wang}
  F.,    {et al.} 2009, \apj, 694, 1643

\bibitem[\protect\citeauthoryear{{Umetsu} \& {Broadhurst}}{{Umetsu} \&
  {Broadhurst}}{2008}]{ume+bro08}
{Umetsu} K.,  {Broadhurst} T.,  2008, \apj, 684, 177

\bibitem[\protect\citeauthoryear{{Younger} \& {Bryan}}{{Younger} \&
  {Bryan}}{2007}]{you+bry07}
{Younger} J.~D.,  {Bryan} G.~L.,  2007, \apj, 666, 647

\bibitem[\protect\citeauthoryear{{Zitrin}, {Broadhurst}, {Barkana}, {Rephaeli}
  \& {Ben{\'{\i}}tez}}{{Zitrin} et~al.}{2011a}]{zit+al11}
{Zitrin} A.,  {Broadhurst} T.,  {Barkana} R.,  {Rephaeli} Y.,
  {Ben{\'{\i}}tez} N.,  2011a, \mnras, 410, 1939

\bibitem[\protect\citeauthoryear{{Zitrin}, {Broadhurst}, {Bartelmann},
  {Rephaeli}, {Oguri}, {Ben{\'{\i}}tez}, {Hao} \& {Umetsu}}{{Zitrin}
  et~al.}{2011b}]{zit+al11b}
{Zitrin} A.,  {Broadhurst} T.,  {Bartelmann} M.,  {Rephaeli} Y.,  {Oguri} M.,
  {Ben{\'{\i}}tez} N.,  {Hao} J.,    {Umetsu} K.,  2011b, ArXiv: 1105.2295

\bibitem[\protect\citeauthoryear{{Zitrin}, {Broadhurst}, {Rephaeli} \&
  {Sadeh}}{{Zitrin} et~al.}{2009}]{zit+al09}
{Zitrin} A.,  {Broadhurst} T.,  {Rephaeli} Y.,    {Sadeh} S.,  2009, \apjl,
  707, L102

\end{thebibliography}

\setlength{\bibhang}{2.0em}

\end{document}